\documentclass[twocolumn]{aastex61}

\pdfoutput=1

\newcommand{\ha}{H$\alpha$}
\newcommand{\oiii}{[{\sc Oiii}]}
\newcommand{\hb}{H$\beta$}
\newcommand{\nii}{[{\sc Nii}]}
\newcommand{\oii}{[{\sc Oii}]}
\newcommand{\neiii}{[Ne{\sc iii}]}

\newcommand{\nbh}{${\rm NB}_H$}
\newcommand{\nbk}{${\rm NB}_K$}

\shorttitle{ISM conditions in the [{\sc Oiii}] emitters at $z>3$}
\shortauthors{Suzuki et al.}
\begin{document}

\defcitealias{KK04}{KK04}
\defcitealias{PP04}{PP04}

\title{The interstellar medium in [OIII]-selected star-forming galaxies at $z\sim3.2$}

\correspondingauthor{Tomoko L. Suzuki}
\email{suzuki.tomoko@nao.ac.jp}

\author{Tomoko L. Suzuki}
\affiliation{National Astronomical Observatory of Japan, Osawa 2-21-1, Mitaka, Tokyo, 181-8588, Japan}
\affiliation{Department of Astronomical Science, SOKENDAI (The Graduate University for Advanced Studies), Osawa 2-21-1, Mitaka, Tokyo, 181-8588, Japan}

\author{Tadayuki Kodama}
\affiliation{Astronomical Institute, Tohoku University, Aramaki, Aoba-ku, Sendai, 980-8578, Japan}

\author{Masato Onodera}
\affiliation{Subaru Telescope, National Astronomical Observatory of Japan, National Institutes of Natural Sciences (NINS), 650 North A’ohoku Place, Hilo, HI 96720, USA}
\affiliation{Department of Astronomical Science, SOKENDAI (The Graduate University for Advanced Studies), Osawa 2-21-1, Mitaka, Tokyo, 181-8588, Japan}

\author{Rhythm Shimakawa}
\affiliation{UCO/Lick Observatory, University of California, 1156 High Street, Santa Cruz, CA 95064, USA}

\author{Masao Hayashi}
\affiliation{National Astronomical Observatory of Japan, Osawa 2-21-1, Mitaka, Tokyo, 181-8588, Japan}

\author{Ken-ichi Tadaki}
\affiliation{National Astronomical Observatory of Japan, Osawa 2-21-1, Mitaka, Tokyo, 181-8588, Japan}

\author{Yusei Koyama}
\affiliation{Subaru Telescope, National Astronomical Observatory of Japan, National Institutes of Natural Sciences (NINS), 650 North A’ohoku Place, Hilo, HI 96720, USA}
\affiliation{Department of Astronomical Science, SOKENDAI (The Graduate University for Advanced Studies), Osawa 2-21-1, Mitaka, Tokyo, 181-8588, Japan}

\author{Ichi Tanaka}
\affiliation{Subaru Telescope, National Astronomical Observatory of Japan, National Institutes of Natural Sciences (NINS), 650 North A’ohoku Place, Hilo, HI 96720, USA}

\author{David Sobral}
\affiliation{Department of Physics, Lancaster University, Lancaster, LA1, 4YB, UK}
\affiliation{Leiden Observatory, Leiden University, PO Box 9513, NL-2300 RA Leiden, the Netherlands}

\author{Ian Smail}
\affiliation{Centre for Extragalactic Astrophysics, Durham University, South Road, Durham DH1 3LE, UK}

\author{Philip N. Best}
\affiliation{SUPA, Institute for Astronomy, Royal Observatory of Edinburgh, Blackford Hill, Edinburgh EH9 3HJ, UK}

\author{Ali A. Khostovan}
\affiliation{Department of Physics and Astronomy, University of California, 900 University Ave., Riverside, CA 92521, USA}

\author{Yosuke Minowa}
\affiliation{Subaru Telescope, National Astronomical Observatory of Japan, National Institutes of Natural Sciences (NINS), 650 North A’ohoku Place, Hilo, HI 96720, USA}
\affiliation{Department of Astronomical Science, SOKENDAI (The Graduate University for Advanced Studies), Osawa 2-21-1, Mitaka, Tokyo, 181-8588, Japan}

\author{Moegi Yamamoto}
\affiliation{Department of Astronomical Science, SOKENDAI (The Graduate University for Advanced Studies), Osawa 2-21-1, Mitaka, Tokyo, 181-8588, Japan}
\affiliation{National Astronomical Observatory of Japan, Osawa 2-21-1, Mitaka, Tokyo, 181-8588, Japan}


\begin{abstract}

We present new results from near-infrared spectroscopy with Keck/MOSFIRE 
of \oiii-selected galaxies at $z\sim3.2$. 
With our $H$ and $K$-band spectra, 
we investigate the interstellar medium (ISM) conditions, 
such as ionization states and gas metallicities. 
\oiii\ emitters at $z\sim3.2$ show 
a typical gas metallicity of 
$\mathrm{12+log(O/H) = 8.07\pm0.07}$ at $\mathrm{log(M_*/M_\odot) \sim 9.0-9.2}$ and 
$\mathrm{12+log(O/H) = 8.31\pm0.04}$ at $\mathrm{log(M_*/M_\odot) \sim 9.7-10.2}$  
when using the empirical calibration method. 
We compare the \oiii\ emitters at $z\sim3.2$ with UV-selected galaxies 
and Ly$\alpha$ emitters at the same epoch 
and find that the \oiii-based selection does not appear to show any systematic bias 
in the selection of star-forming galaxies. 
Moreover, comparing with star-forming galaxies at $z\sim2$ 
from literature, 
our samples show similar ionization parameters and gas metallicities as those obtained 
by the previous studies using the same calibration method. 
We find no strong redshift evolution in the ISM conditions between $z\sim3.2$ and $z\sim2$. 
Considering that the star formation rates at a fixed stellar mass also do not significantly change 
between the two epochs, 
our results support the idea that the stellar mass is the primary quantity to describe 
the evolutionary stages of individual galaxies at $z>2$. 
\end{abstract}




\section{Introduction} \label{sec:intro}

Recent near-infrared (NIR) spectroscopic surveys have suggested 
that star-forming galaxies at high redshifts ($z>1$) 
typically have different interstellar medium (ISM) conditions 
from those found in local star-forming galaxies 
\citep[e.g.][]{masters14,steidel14,hayashi15,shapley15,holden16,kashino17}. 
Star-forming galaxies at high redshifts 
show a systematic offset from local galaxies 
on the Baldwin-Phillips-Terlevich diagram (so called BPT diagram; 
\citealt{bpt,veilleux87}), 
i.e. they have higher \oiii/\hb\ ratios with respect to \nii/\ha\ 
\citep[e.g.][]{erb06-644,masters14,steidel14,shapley15,kashino17}. 
Also on a stellar mass versus \oiii/\hb\ ratio diagram (Mass--Excitation diagram; \citealt{juneau11}), 
star-forming galaxies at high redshifts 
show systematically higher \oiii/\hb\ ratios than local ones  
at a fixed stellar mass \citep[e.g.][]{cullen14,shimakawa15a,holden16,strom16,kashino17}. 
These differences suggest 
that ISM conditions at high redshifts are different 
as a result of lower gas metallicities, 
higher ionization parameters, harder spectra of ionizing sources, 
and the combination of all these factors 
\citep[e.g.][]{kewley13,nakajima14,steidel14,steidel16,trainor16,strom16,kashino17}. 

The relation between stellar mass and gas metallicity of star-forming galaxies 
has been investigated by several studies. 
It has been known that there is a positive correlation between 
stellar mass and gas metallicity since about 40 years ago \citep{lequeux79}.  
Now the stellar mass--gas metallicity relation is observed 
for star-forming galaxies from $z=0$ even up to $z\sim5$ 
\citep{tremonti04,erb06-644,maiolino08,mannucci09,henry13,stott13,cullen14,steidel14,troncoso14,wuyts14,yabe15PASJ,zahid14,sanders15,faisst16,onodera16}, 
and star-forming galaxies at higher redshifts have lower gas metallicities 
than local star-forming galaxies at a fixed stellar mass. 

When estimating the gas metallicities of star-forming galaxies, 
the strong line methods are often used. 
The relations between strong emission line ratios 
and gas metallicities are obtained empirically using local star-forming galaxies 
(e.g. \citealt{PP04,maiolino08,curti16}, and at $z=0.8$ by \citealt{jones15}) 
or with the photoionization models 
\citep[e.g.][]{kewleydopita02}. 
It has been suggested that, however, the locally calibrated relations 
are no longer applicable to star-forming galaxies at high redshifts 
because the typical ISM conditions of star-forming galaxies 
seem to change from $z=0$ to higher redshifts 
\citep[e.g.][]{kewley13,nakajima14,steidel14,kashino17}. 
It is still under discussion whether we can adopt the locally calibrated methods 
to star-forming galaxies at higher redshifts 
because some studies have reported that the physical conditions of 
{\sc Hii} regions do not evolve with redshifts at a fixed metallicity 
\citep[e.g.][]{jones15,sanders16}. 
Moreover, 
it is known that the gas metallicities calibrated with different emission line ratios 
show systematic offsets from one another \citep{kewleyellison08}.

Studies of the ISM conditions and the mass--metallicity relation 
mainly target star-forming galaxies at $z<$ 2--2.5, 
up to the highest peak of galaxy formation and evolution \citep[e.g.][]{hopkins06,madau14,khostovan15}. 
However, 
the epoch of $z>3$ is also important 
because the cosmological inflow is likely to be prominent at this epoch \citep[e.g.][]{mannucci09,cresci10,troncoso14}. 
The gas-phase metallicity of a galaxy reflects the relative contributions from star formation, 
gas outflow and gas inflow. 
Therefore, 
the metal content of galaxies is one of the key quantities 
in order to reveal how the gas inflow/outflow processes, as well as star formation,  
have an impact on galaxy formation and evolution.

NIR spectroscopic observations of star-forming galaxies at $z>3$ 
have been carried out by targeting UV-selected galaxies, 
such as Lyman break galaxies (LBGs) and Ly$\alpha$ emitters 
\citep[LAEs; e.g.][]{steidel96,steidel03,maiolino08,mannucci09,troncoso14,holden16,onodera16,nakajima16}.
However, 
the evolution of the ISM conditions and the mass--metallicity relation especially at $z>3$
has not yet been fully understood  
because of the large uncertainties related to the estimation of gas metallicities and 
the limited sample sizes at this epoch \citep[e.g.][]{onodera16}. 
Additionally, 
at $z>3$, 
it is difficult to obtain a representative sample of star-forming galaxies 
because available indicators of star-forming galaxies are limited. 
Since the UV-selected galaxies tend to be biased towards less dusty galaxies \citep{oteo15}, 
it is important to obtain a sample of star-forming galaxies using other selection techniques,  
which are less affected by dust extinction than the UV light. 
Rest-frame optical emission lines are very useful for this purpose.

There are some methods to select galaxies based on the strength of emission lines. 
The grism spectroscopy at the $H$-band by the {\it Hubble Space Telescope} {\it (HST)} can 
pick up galaxies at $z\sim$1--3 with strong emission lines 
in the rest-frame optical \citep[e.g.][]{3dhst}. 
\citet{maseda13,maseda14} selected 
extreme emission line galaxies at $z\sim$1-2 based on the emission line flux 
and equivalent width from the {\it HST} NIR grism spectroscopy. 
Their sample consists of low mass galaxies of $\rm log(M_*/M_\odot)\sim8-9$.
They showed that the extreme emission line galaxies are in the starburst phase with 
high specific star-formation rates (SFRs) 
and have high \oiii/\hb\ ratios ($\ge5$). 
\citet{hagen16} also used the {\it HST} NIR grism data 
to construct a sample of the optical emission line-selected galaxies at $z\sim2$. 
Comparing the sample with LAEs at similar redshifts, 
they found that the two galaxy populations have similar physical quantities 
in a stellar mass range of $\rm log(M_*/M_\odot)\sim7.5-10.5$.

Imaging observations with a narrow-band (NB) filter 
are also a very efficient way of constructing a sample of emission line galaxies 
in a particular narrow redshift slice 
\citep[e.g.][]{bunker95,teplitz99,moorwood00,geach08,sobral13,tadaki13}.  
At $z>3$, 
the \ha\ emission line, 
which is one of the most reliable tracers of star-forming galaxies, 
is no longer accessible from the ground. 
We need to use other emission lines at shorter wavelengths,  
such as \oiii, \hb, and \oii\ \citep{khostovan15,khostovan16}. 
As mentioned above, 
normal star-forming galaxies at high redshifts tend to show 
brighter \oiii\ emission lines. 
While there is a clear trend of decreasing \oiii/\hb\ ratio with 
increasing stellar mass \citep{juneau11,juneau14,strom16}, 
the \oiii\ emission lines would be observable even for massive star-forming galaxies 
at $z>3$ because they are bright in \oiii\ intrinsically.

Is the \oiii\ emission line a useful tracer of star-forming galaxies 
at higher redshifts actually? 
\citet{suzuki15} have found that the \oiii-selected galaxies at $z>3$ 
show a positive correlation between stellar mass and SFR, 
which is known as the ``main-sequence'' of star-forming galaxies 
\citep[e.g.][]{whitaker12,kashino13,tomczak16}. 
This suggests that we can trace the typical star-forming galaxies at $z>3$ 
using the \oiii\ emission line. 
Moreover, \citet{suzuki16} 
have shown that the \oiii-selected galaxies 
show similar distributions of stellar mass, SFR, and dust extinction 
as those of normal \ha-selected star-forming galaxies at $z\sim2.2$, 
supporting the idea that the \oiii\ emission line can be used as a tracer 
of star-forming galaxies at high redshifts. 
%%%
Therefore, 
the \oiii-selected galaxies can probe dustier star-forming galaxies 
which are likely to be missed by the UV-based or \oii\ selection \citep{hayashi13}. 
%%%
We also note that 
another great advantages of NB-selected galaxies 
is the high efficiency of follow-up observations 
because their line fluxes and redshifts are obtained in advance 
by the NB imaging observations.

In this paper, 
we present the results obtained from the spectroscopic observation 
of \oiii\ emitters at $z=3.24$ in the COSMOS field obtained 
by the HiZELS survey \citep{geach08,sobral09,sobral13,best13,khostovan15}. 
We carried out $H$ and $K$-band spectroscopy  
of the \oiii\ emitters with Keck/MOSFIRE.  
We investigate  
the physical conditions of the \oiii\ emitters at $z>3$ 
such as their ionization states and gas metallicities. 

This paper is organized as follows: 
in Section \ref{sec:target}, 
we present our parent sample of \oiii\ emitters at $z\sim3.2$. 
We also describe our NIR spectroscopy of the \oiii\ emitters 
with Keck/MOSFIRE, 
and the details of the observations and data reduction/analyses. 
In Section \ref{sec:z3}, 
we show our results about the ISM conditions of our sample, 
and compare with other galaxy populations at the same epoch. 
In Section \ref{sec:comparewithz2}, 
we discuss the evolution of star-forming activities and ISM conditions of star-forming galaxies 
between $z\sim3.2$ and $z\sim2.2$. 
Finally we summarize this work in Section \ref{sec:summary}. 

Throughout this paper, 
we assume the cosmological parameters of 
$\Omega _\mathrm{m}=0.3$, $\Omega_\mathrm{\Lambda}=0.7$, 
and $H_\mathrm{0}=70\ \mathrm{km\ s^{-1} Mpc^{-1}}$. 
All the magnitudes are given in AB system, 
and we adopt the Chabrier initial mass function (IMF; \citealt{chabrier03}) 
unless otherwise noted. 
We refer to wavelengths of all emission lines using vacuum wavelengths.

\section{Sample selection, observations, and reduction}\label{sec:target}

\subsection{Selection of [OIII] candidate emitters at $z\sim3.24$}\label{subsec:hizels}

HiZELS (the High-$z$ Emission Line Survey; \citealt{sobral12,sobral13}, see also \citealt{best13}) 
is a systematic NB imaging survey 
using NB filters in the $J$, $H$, and $K$-bands of the Wide Field CAMera (WFCAM; \citealt{casali07}) 
on the United Kingdom Infrared Telescope (UKIRT), 
and the NB921 filter of the Suprime-Cam \citep{miyazaki02} on the Subaru Telescope.   
Emission line galaxy samples used in this study are based on the HiZELS catalogue 
in the Cosmological Evolution Survey (COSMOS; \citealt{scoville07}) field. 

With the $\mathrm{H_{2}S1}$ filter (hereafter \nbk\, $\lambda_\mathrm{c} = 2.121\ \mu \mathrm{m}$, and $\mathrm{FWHM = 210\ \AA}$)
of WFCAM, 
HiZELS selects the \oiii$\lambda$5008 emission from galaxies at $z=3.235 \pm 0.021$. 
Here we construct a catalog of \oiii\ emitters at $z\sim3.24$  
by combining 
the \nbk\ emitter catalog from HiZELS \citep{sobral13} 
and the latest photometric catalog in the COSMOS field 
(COSMOS2015; \citealt{cosmos2015}) 
in a similar way to \citet{khostovan15}.   
The COSMOS2015 catalog includes 
the new deep NIR and IR data from UltraVISTA-DR2 survey and 
from the SPLASH ({\it Spitzer} Large Area Survey with Hyper-Suprime-Cam) project \citep{cosmos2015}. 
Such deep IR photometry becomes more important 
when estimating photometric redshifts and stellar masses of galaxies at higher redshifts.

In the first place, we search for counterparts of the \nbk\ emitters 
in the COSMOS2015 catalog 
with a searching radius of $0''.6$. 
The selection of the NB emitters are based on the color excess of NB 
with respect to broad band (BB), 
and the equivalent width. 
A parameter $\Sigma$ is introduced to quantify the significance of a NB excess 
relative to 1$\sigma$ photometric error \citep{bunker95}. 
This parameter $\Sigma$ is represented as a function of NB magnitude as follows 
\citep{sobral13}: 

\begin{equation}
\Sigma = \frac{1 - 10^{-0.4(K-{\rm{NB})}}}{10^{-0.4({\rm ZP-NB})} \sqrt{\pi r^2_{\rm ap} (\sigma ^2_{\rm NB} + \sigma ^2_{K})}}, 
\end{equation}

\noindent
where NB and BB are NB and BB magnitudes, 
ZP is the zero-point of the NB 
(the BB images are scaled to have the same ZP as the NB images), 
$r_\mathrm{ap}$ is the aperture radius in pixel, 
and $\sigma _\mathrm{NB}$ and $\sigma _\mathrm{BB}$ are 
the rms per pixel of the NB and BB images, respectively \citep{sobral13}. 
Emission line fluxes, $F_\mathrm{line}$, and the rest-frame equivalent widths, $\mathrm{EW_{rest}}$, are calculated with 

\begin{equation}
F_\mathrm{line} = \Delta_\mathrm{NB} \frac{f_\mathrm{NB} - f_\mathrm{BB}}{1-\Delta _\mathrm{NB}/\Delta _\mathrm{BB}}, 
\end{equation}

\noindent
and 

\begin{equation}
\mathrm{EW_{rest}} = \Delta_\mathrm{NB} \frac{f_\mathrm{NB} - f_\mathrm{BB}}{f_\mathrm{BB} - f_\mathrm{NB}(\Delta_\mathrm{NB}/\Delta_\mathrm{BB})}, 
\end{equation}

\noindent
where $f_\mathrm{NB}$ and $f_\mathrm{BB}$ are the flux densities for NB and BB, 
and $\Delta_\mathrm{NB}$ and $\Delta_\mathrm{BB}$ are the FWHMs of the NB and BB filters, respectively \citep[e.g.][]{tadaki13}. 
The selection criteria of the NB emitters are $\Sigma>3$ and 
the observed-frame equivalent-width of $\mathrm{EW_{obs}} \ge 80.8\ \AA$  
(the rest-frame EW $\sim$ 19~\AA\ for \oiii\ at $z=$3.24, \citealt{sobral13,khostovan15}).  
We select \oiii\ candidate emitters at $z\sim3.24$ 
with photometric redshifts of $2.8 < z_\mathrm{photo} < 4.0$. 
Additionally, 
we employed color--color diagrams 
($UVz$ and $Viz$) 
for the emitters with no photometric redshifts 
in the COSMOS2015 catalog following the methods 
introduced in  \citet{khostovan15}. 
We finally obtained 174 \oiii\ candidate emitters at $z\sim3.24$ in the COSMOS field.

%% Observation 
\subsection{$H$ and $K$-band spectroscopy with Keck/MOSFIRE}\label{subsec:obs}

Observations were carried out on the first half night on 27th March 2016 
with the Multi-Object Spectrometer For Infra-Red Exploration 
(MOSFIRE; \citealt{mclean10,mclean12}) 
on the Keck~I telescope as a Subaru-Keck time exchange program (S16A-058; PI: T. Suzuki). 
The wavelength resolution of MOSFIRE is $R\sim3600$. 
Slit widths were set to be $0.7''$. 
Our primary targets are ten \oiii\ candidate emitters at $z\sim3.24$, 
which are chosen so that we can maximize the number of \oiii\ emitters 
in one MOSFIRE pointing. 
We filled the unused mask space with ten photometric redshift-selected sources 
with $K<24~{\rm mag}$ at $3.0 < z_{\rm photo} < 3.5$.  
We obtained their spectra in $K$ and $H$-bands in order to detect 
the major emission lines, such as \oiii$\lambda \lambda$5008,4960, \hb, and \oii$\lambda \lambda$3727,3730.  
The total integration time was 120 min and 90 min for $K$ and $H$-band, respectively. 
The seeing (FWHM) was $0.7''$--$1.0''$.

%% Reduction 
\subsection{Data reduction and analyses}\label{subsec:reduction}

The obtained raw spectra were reduced using the MOSFIRE Data Reduction 
Pipeline\footnote{\url{https://keck-datareductionpipelines.github.io/MosfireDRP/}} ({\tt MosfireDRP}), 
which is described in more detail in \citet{steidel14}. 
The pipeline follows the standard data reduction procedures: 
flat-fielding, wavelength calibration, sky subtraction, rectification, and combining the individual frames. 
Finally we obtained the rectified two-dimensional (2D) spectra. 
One-dimensional (1D) spectra were extracted from the 2D spectra 
with $1.3''$--$1.8''$ diameter aperture in order to maximize the signal-to-noise (S/N) ratio.  
The telluric correction and flux calibration 
were carried out by using a standard A0V star, HIP43018, 
which were taken at the same night.

%% emission line measurements 
All of the ten NB-selected \oiii\ candidate emitters clearly show the \oiii\ doublet lines in the $K$-band (100\% detection), 
and are identified as \oiii\ emitters at $z=$ 3.23--3.27.    
Our observations demonstrate  
the extremely low contamination of the NB-selected galaxies \citep{sobral13,khostovan15} 
and also the high efficiency of follow-up observations.  
The \hb\ and \oii\ emission lines are also visually identified in the 1D spectra 
in the $K$- and $H$-band, respectively, 
for all of the \oiii\ emitters.  
As for the photometric redshift-selected targets,   
seven sources are identified as the galaxies at $z=$3.00-3.45 with their \oiii\ doublets 
yielding a 70\% detection.

%% Flux calibration 
We included a monitoring star in our mask so that we can use it 
to correct for different seeing conditions when observing  
the science targets and the standard star. 
By comparing the observed fluxes of the star 
 with the 2MASS magnitudes, 
we determine the correction factors of 
$1.22\pm0.04$ and $0.89\pm0.03$ for $H$ and $K$-band, respectively. 
%%%
We note that we have corrected for 
the slit loss by using the standard star and the monitoring star, 
if the sources are well approximated by the point sources. 
Even if the sources are extended, 
slit losses would be not very important here 
because our analysis is not strongly depend on absolute fluxes.

\begin{figure}[tp]
\centering\includegraphics[width=1.0\columnwidth]{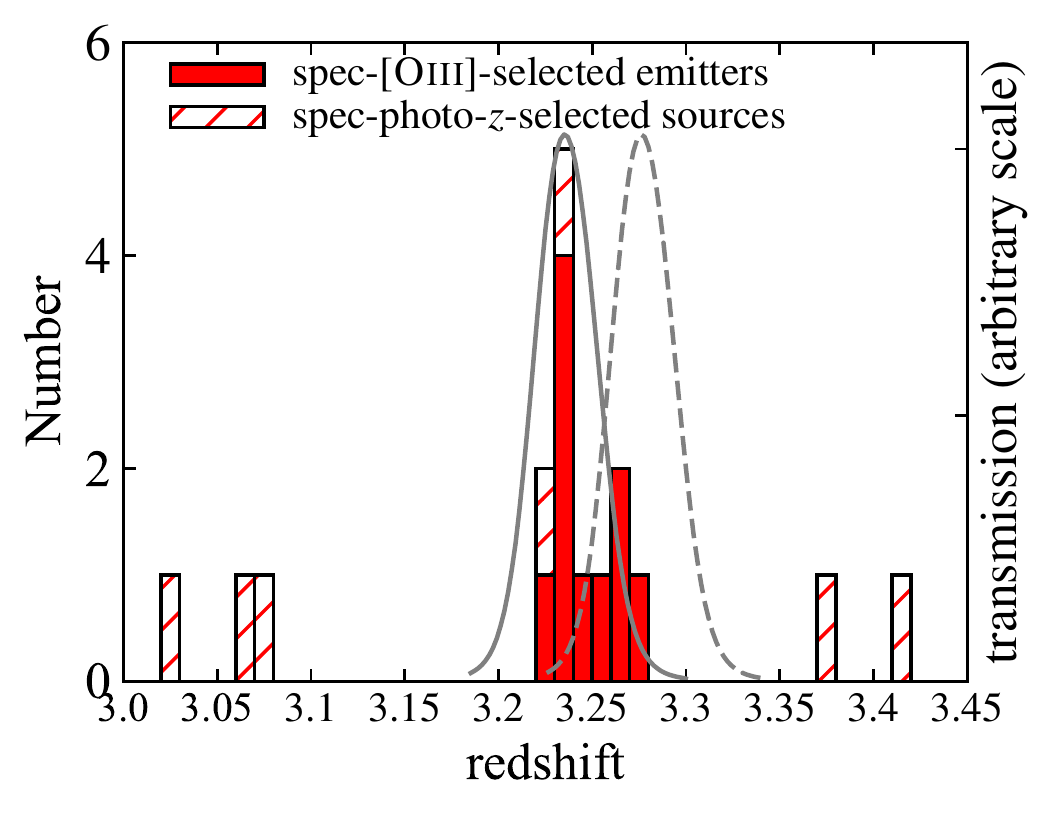}
\caption{Redshift distribution of the spectroscopically confirmed sources from this observation.  
 The filled histogram shows the \oiii\ emitters 
 and the hatched histogram shows that of our secondary targets, 
 i.e. the photo-$z$-selected sources. 
 The transmission curves of the \nbk\ filter are also shown. 
 The wavelength range of the \nbk\ filter is converted to the redshift ranges 
for the \oiii$\lambda$5008 emission line (the solid curve) and 
the \oiii$\lambda$4960 emission line (the dashed curve), respectively.}
\label{fig:specz}
\end{figure}

In order to measure the emission line fluxes, 
we perform Gaussian fitting for the emission lines 
using the {\tt SPECFIT} 
\footnote{\url{http://stsdas.stsci.edu/cgi-bin/gethelp.cgi?specfit}}
\citep{kriss94} in {\tt STSDAS} of the {\tt IRAF} environment. 
%%% 
At first, we fit the \oiii\ doublet and \hb\ 
with a Gaussian by assuming a common velocity dispersion. 
The \oiii\ doublet lines are fitted 
by assuming the line ratio \oiii$\lambda$5008/\oiii$\lambda$4960 of 3.0 \citep{storey00}. 
Redshifts of the sources are determined 
using the \oiii\ line at 5008.24~\AA.  
The redshift distribution of our sample is 
shown in Figure~\ref{fig:specz}. 
Then, the \hb\ line and \oii\ doublet lines are fitted assuming the determined redshifts and velocity dispersions. 
We also fit relatively weak lines, 
such as He{\sc ii}$\lambda$4687 and [Ne{\sc iii}]$\lambda$3870, 
by assuming the determined redshifts and velocity dispersions. 
Errors of the fitted line fluxes 
are obtained by taking into account 
the wavelength-dependent sky noise due to the O/H sky lines 
and the errors from $\chi^2$ fitting.

For all of the \oiii\ emitters, 
the \oiii$\lambda$5008 lines are detected with very high S/N ratios, $\mathrm{S/N}>20$.
The \hb\ line is also detected for all the emitters at more than 3$\sigma$ significance levels. 
Although there are some cases of the \oii$\lambda$3727 doublet lines being affected 
by OH skylines, the summed flux of the doublet lines is detected at more than 3$\sigma$ 
levels for all the emitters. 
As for the \neiii\ emission line, 
it is detected from six emitters at more than 3$\sigma$ significance levels. 
The He{\sc ii} line is not detected at $\mathrm{S/N} > 3$ for any of the \oiii\ emitters. 
For the photo-$z$-selected sources, 
the \oiii$\lambda$5008 and the summed \oii$\lambda$3727 fluxes 
are detected at more than 3$\sigma$ significance levels. 
For some sources, 
their \hb\ or \neiii\ emission lines overlap with OH skylines.  
We find that two of the photo-$z$-selected sources, which are within the redshift coverage 
of the \nbk\ filter, are not selected as the emitters due to their relatively weak \oiii$\lambda$5008 fluxes. 
The reduced spectra and estimated fluxes are shown in Appendix-\ref{sec:spectra} all together.

The velocity dispersions obtained by the emission line fitting for each galaxy 
yield values of 140--310 ${\rm km\ s^{-1}}$ in the rest-frame. 
From the fact that all \hb\ lines are narrow ($\ll 1000 \mathrm{km\ s^{-1}}$), 
we consider that there is no obvious broad-line AGN in our sample. 
We also note that none of our sources is detected at X-ray with {\it Chandra} \citep{civano16}.

The redshift distribution of our sample is shown in Figure \ref{fig:specz}.  
We find that three \oiii\ emitters are located at slightly higher redshifts  
than the redshift range expected for the \oiii$\lambda$5008 line with the \nbk\ filter.  
In Figure \ref{fig:specz}, 
we show the transmission curves of the \nbk\ filter as a function of redshift 
in the two cases; one for the \oiii$\lambda$5008 line 
and the other for the \oiii$\lambda$4960 line.  
The three \oiii\ emitters at slightly higher redshifts 
turn out to be detected by their strong \oiii$\lambda$4960 with the \nbk\ filter.  
The fraction of the \oiii$\lambda$4960 emitters is $\sim$ 30 \%, 
and this is consistent with our estimation from the luminosity function at $z=2.23$
in \citet{suzuki16} and the result of the spectroscopy of \oiii+\hb\ emitters at $z=1.47$ 
by \citet{sobral15}. 
\hb\ emitters are not found in our target sample.

\subsection{Stellar absorption correction for H$\beta$}\label{subsec:hbcorrection}
In the following analyses,
we use the \hb\ fluxes corrected for the stellar absorption. 
We assume the typical EW of the absorption line of 2~\AA\ \citep{nakamura04}, 
and use the continua estimated from the $Ks$-band magnitudes  
after subtracting the contributions from emission lines. 
The stellar-absorption-corrected \hb\ fluxes are 
estimated by  

\begin{equation}
F_{\rm H\beta, corr} = F_{\rm H\beta, obs} + 2~{\rm (\AA)} \times (1+z) \times f_c, 
\end{equation}

\noindent
where $f_c$ is a continuum flux density. 
The correction factors for the \hb\ stellar absorption ($F_{\rm H\beta, corr}/F_{\rm H\beta, obs}$) 
are $\sim$1.0--1.2.

%%%
\subsection{Estimation of physical quantities}\label{sec:estimation}

The stellar masses of the spectroscopically confirmed sources 
are estimated by SED fitting with the public code 
 {\tt EAZY} \citep{brammer08} and {\tt FAST} \citep{kriek09b}. 
 We use the total magnitudes of 14 photometric bands; 
$u, B, V, r, i', z'', Y, J, H, K_s$, 3.6, 4.5, 5.8, and 8.0 $\mu$m 
from the COSMOS2015 catalog. 
We subtract the contributions of the emission lines, the \oiii\ doublet and \hb, and \oii\ doublet, 
from the $Ks$ and $H$-band magnitudes, respectively, before the SED fitting. 
When running the {\tt FAST}, we fix their redshifts to those measured from the spectroscopy. 
We use the population synthesis models of \citet{bc03} with a Chabrier IMF \citep{chabrier03},  
and the dust extinction law of \citet{calzetti00}. 
We assume exponentially declining SFHs with ${\rm log(\tau/yr)=}$ 8.5--11.0 
in steps of 0.1, and metallicities of $Z=0.004, 0.008,$ and 0.02 (solar). 

SFRs are estimated from UV continuum luminosities 
in order to compare with a whole sample of \oiii\ candidate emitters (Figure~\ref{fig:SFMS_all}).  
Dust extinction is corrected for using the slope of the rest-frame
UV continuum spectrum (e.g. \citealt{meurer99,heinis13}).
The UV slope $\beta$ is defined as $f_{\lambda}$$\propto$$\lambda^{\beta}$.
We estimate $\beta$ by fitting a linear function
to the five broad-bands from the $B$ to $i$-band. 
The slope $\beta$ is converted to dust extinction $A_{\rm FUV}$ with the following equation from \citet{heinis13}:

\begin{equation}
A_{\rm FUV} = 3.4 + 1.6 \beta.
\label{eq:beta}
\end{equation}

\noindent 
Then, the intrinsic flux density $f_{\nu, {\rm int}}$ is obtained from 

\begin{equation}
f_{\nu, {\rm int}} = f_{\nu, {\rm obs}}\ 10^{0.4A_{\rm FUV}}.  
\end{equation}

${\rm SFR_{UV}}$ is estimated from the $r$-band 
($\lambda _{\rm c} = 6288.7\ \mathrm{\AA}$ which corresponds to $\lambda_\mathrm{0} = 1500 \mathrm{\AA}$ at $z=3.2$) magnitude using the equation from \citet{madau98}:

\begin{eqnarray}\nonumber
{\rm SFR \ (M_\odot yr^{-1})} &=& \frac{4\pi D_L^2  f_{\nu, {\rm int}}}{(1+z) \times 8 \times 10^{27} \ ({\rm erg \ s^{-1} cm^{-2}  Hz^{-1}})} \\
&=& \frac{L({1600} {\rm \AA})}{8 \times 10^{27} \ ({\rm erg \ s^{-1} Hz^{-1}})}, 
\label{eq:SFR_uv}
\end{eqnarray}

\noindent
where $D_L$ is the luminosity distance.  
Considering the difference between Chabrier and Salpeter \citep{salpeter55} IMFs, 
we divide the SFRs by a factor of 1.7 \citep{pozzetti07} 
so that we always use Chabrier IMF throughout this paper.

For the two photo-$z$-selected sources, 
which are not included in the COSMOS2015 catalog,  
we use the photometric data 
($u, B, V, g, r, i, z, J, K$) 
from the catalog of \citet{ilbert09}. 
The estimated stellar mass, dust extinction, and $\mathrm{SFR_{UV}}$ 
for each galaxy are summarized in Appendix-\ref{sec:spectra}.

Comparing the estimated $\rm SFR_{UV}$ 
with those obtained by {\tt FAST}, 
the results of the SED fitting show a systematic offset of $\sim+0.25$~dex 
with respects to those obtained from the rest-frame UV luminosities. 
Since we compare SFRs obtained with the same method in Section~\ref{sec:SFMS}, 
such a systematic offset does not affect our results. 
As for $A_{\rm FUV}$, 
there is no systematic offset and differences between the two methods 
are within 0.4~mag.

%%%
In addition to $\mathrm{SFR_{UV}}$, 
we also estimate SFRs from the \hb\ luminosities. 
The dust extinction for \hb\ is corrected for by using the UV slope $\beta$ \citep{heinis13}, 
and the Calzetti extinction law \citep{calzetti00} 
assuming $E(B-V)_{\rm nebular}=E(B-V)_{\rm stellar}$ 
\citep[e.g.][]{erb06-647,reddy10,reddy15}. 
We convert the dust-extinction-corrected \hb\ luminosity to the \ha\ luminosity  
using the intrinsic \ha/\hb\ ratio of 2.86 under the assumption of Case B recombination 
with a gas temperature $T_e=10^4$ K and an electron density $n_e=10^2\ {\rm cm^{-3}}$ \citep{osterbrock06}.  

Then we convert the estimated \ha\ luminosities to SFRs 
using the equation from \citet{kennicuttevans12};

\begin{equation}
\mathrm{log(SFR_{H\alpha}/M_\odot yr^{-1})} = \mathrm{log}(L_\mathrm{H\alpha}/\mathrm{erg\ s^{-1}}) - 41.27. 
\end{equation}

\noindent 
Here we account for the difference between the Chabrier and Kroupa IMF  
by subtracting 0.013~dex \citep{pozzetti07,marchesini09}.

\begin{figure}
\centering\includegraphics[width=0.8\columnwidth]{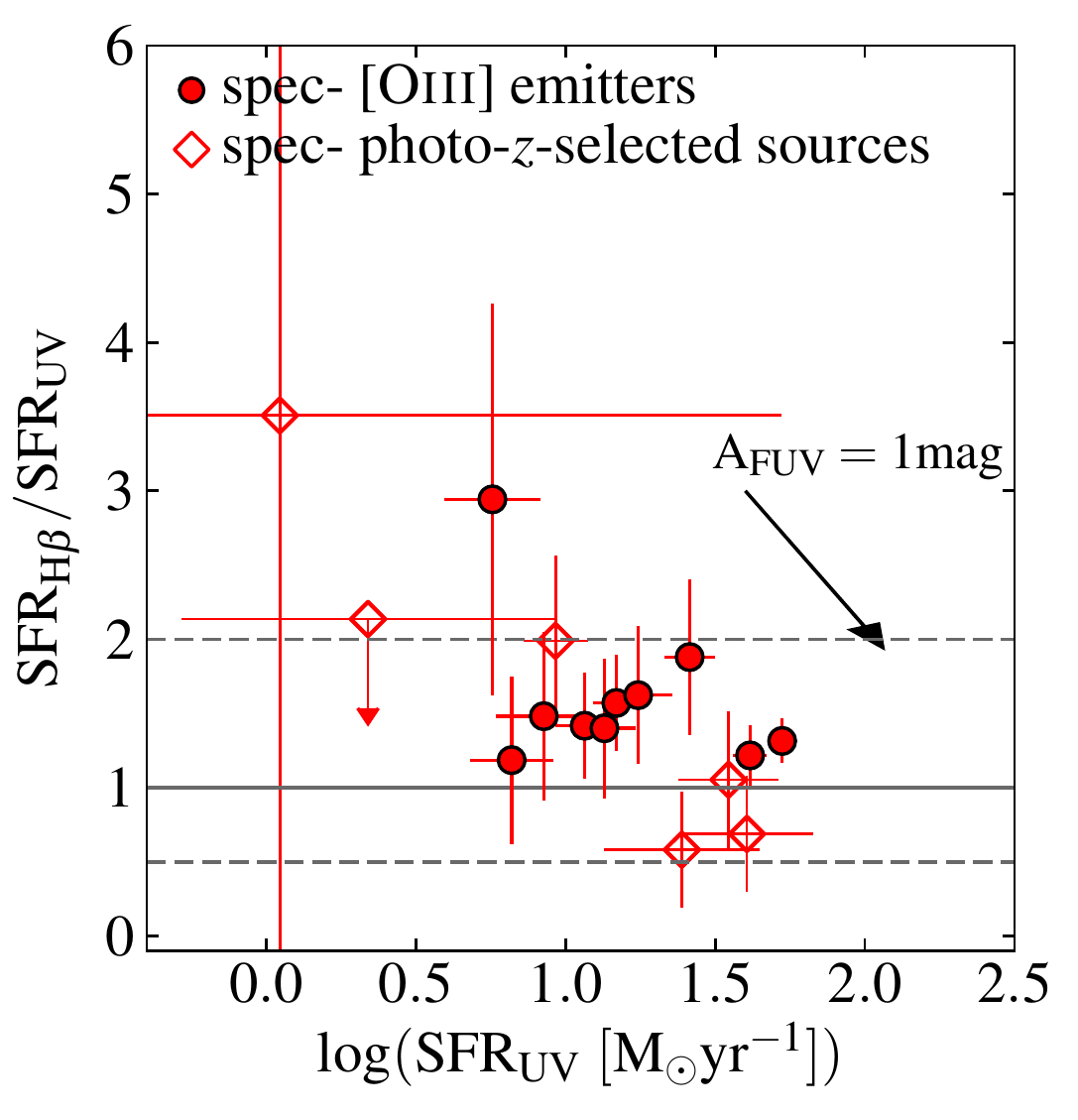}
\caption{
$\mathrm{SFR_{UV}}$ versus $\mathrm{SFR_{H\beta}/SFR_{UV}}$ ratio 
of our spectroscopically confirmed galaxies. 
Here we do not consider the extra extinction to the nebular emission, 
i.e. we assume $E(B-V)_{\rm nebular}=E(B-V)_{\rm stellar}$ \citep[e.g.][]{erb06-647,reddy10,reddy15}. 
Dust extinction is corrected for by using the UV slope $\beta$ (Eq.~\ref{eq:beta}). 
The solid line represents the case where the two SFRs are identical, 
and the dashed lines represent the cases where the difference between the two is a factor of two. 
The arrow shows how dust correction with $\mathrm{A_{FUV} = 1~mag}$ moves 
the points on this diagram. 
For most of our targets, 
SFRs derived from the two different indicators are identical with each other 
within a factor of two. 
}
\label{fig:SFRcomparison}
\end{figure}

In Figure \ref{fig:SFRcomparison}, 
we compare  the two SFRs derived from UV and \hb\ luminosities.  
We find that 
the two SFRs derived from UV luminosities and from \hb\ luminosities 
have similar values within a factor of two except for a few sources. 
The mean $\rm SFR_{H\beta}/SFR_{UV}$ for our sample 
is $1.6\pm0.2$. 
We can estimate their SFRs reasonably well from the UV luminosities with dust correction 
based on the UV slope at $z>3$.

\subsection{Stellar mass--SFR relation}\label{sec:SFMS}

\begin{figure}[tp]
\centering\includegraphics[width=1.0\columnwidth]{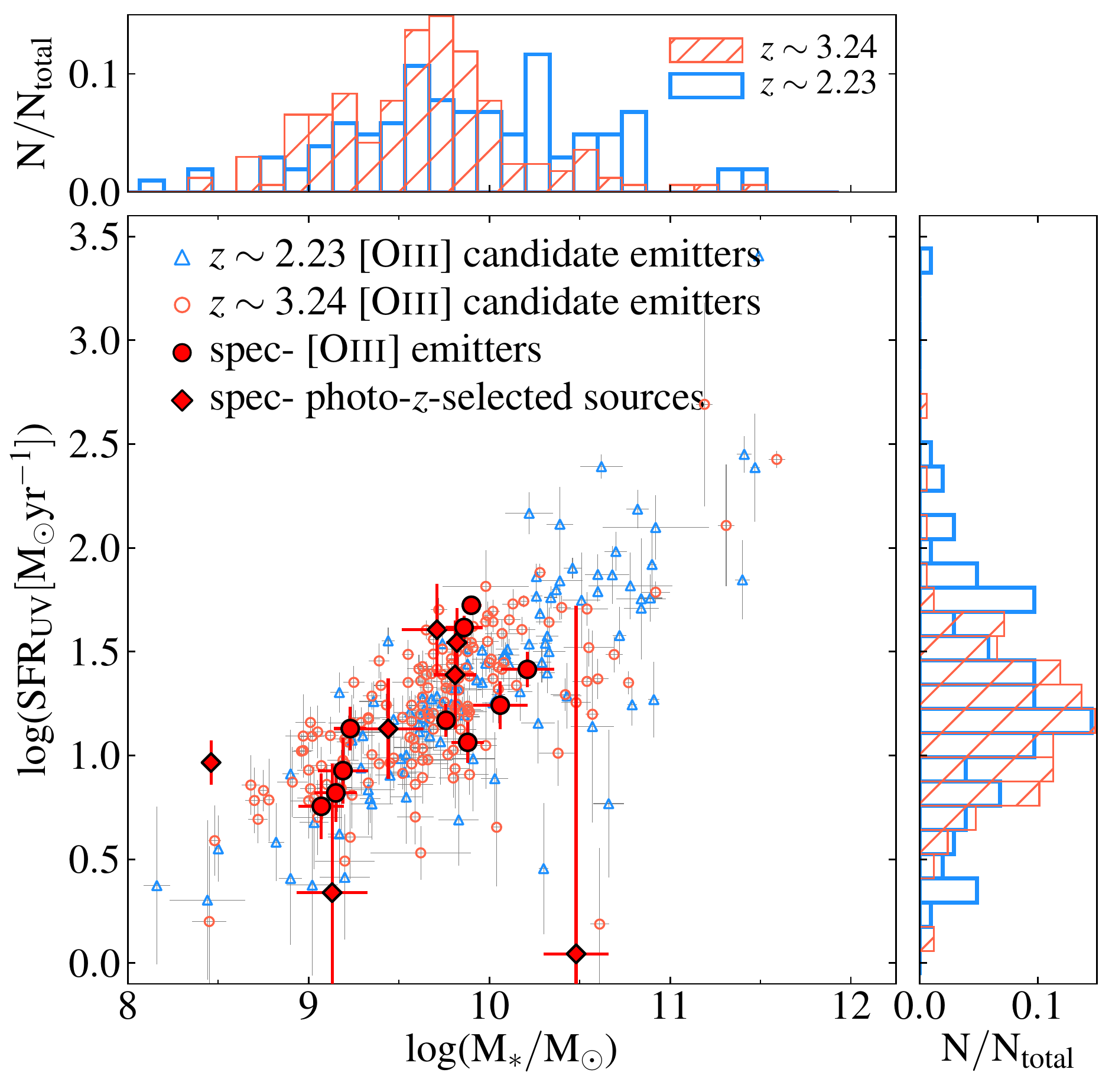}
\caption{Relation between stellar mass and $\mathrm{SFR_{UV}}$. 
The spectroscopically confirmed galaxies in this study are identified. 
\oiii\ candidate emitters at $z\sim3.24$ (open circles) and $z\sim2.23$ (open triangles) 
in the COSMOS field 
are also shown. 
Top and right histogram shows the stellar mass and SFR distribution, respectively. 
Hatched and open histograms correspond the \oiii\ candidate emitters at $z\sim3.24$ and $z\sim2.23$, respectively. 
The spectroscopically confirmed \oiii\ emitters 
are not biased towards a particular region on the stellar mass--$\mathrm{SFR_{UV}}$ 
plane with respect to the parent sample at $z\sim3.24$. 
}
\label{fig:SFMS_all} 
\end{figure}

In Figure~\ref{fig:SFMS_all} we show 
the relation between the stellar masses and $\mathrm{SFR_{UV}}$ of 
the spectroscopically confirmed galaxies in this studies 
together with 
the \oiii\ candidate emitters at $z\sim3.24$ from HiZELS.  
This figure shows that our targets 
are not biased towards a particular region on the stellar mass--$\mathrm{SFR_{UV}}$ diagram 
with respect to the parent sample of the \oiii\ emitters at $z\sim3.24$. 
This indicates that they are {\it normal} star-forming galaxies at the epoch.

We also show the \oiii\ candidate emitters at $z\sim2.23$ 
after matching the \nbh\ emitter catalog in the COSMOS field from HiZELS 
 \citep{sobral13} with the COSMOS2015 catalog. 
 The selection criteria of the \nbh\ emitters are the same as those mentioned in Section~\ref{subsec:hizels} 
 with the \nbh\ filter being used instead of the \nbk \citep{sobral13}. 
 We select \oiii\ candidate emitters at $z\sim2.23$ 
 with photometric redshifts of $1.7 < z_{\rm photo} < 2.8$. 
 We also employ the color--color diagrams ($BzK$, $izK$, and $UVz$)
 for the emitters with no photometric redshifts 
 as introduced in \citet{khostovan15}. 
 We obtained 117 \oiii\ candidate emitters at $z\sim2.23$ in total.

 Stellar masses and $\rm SFR_{UV}$ of the \oiii\ candidate emitters at $z\sim3.24$ and $z\sim$2.23 
 are estimated following the same procedure as described in Section~\ref{sec:estimation}. 
As for \oiii\ emitters at $z\sim2.23$, 
we use the $V$-band magnitude to estimate $\rm SFR_{UV}$. 
 The redshift is fixed of each source is fixed to $z=3.24$ or 2.23. 
We note that 
we take into account the different luminosity limit of the \oiii\ emission line 
when comparing the \oiii\ emitters at different redshifts in Figure~\ref{fig:SFMS_all}.

We find that the \oiii\ emitters at $z\sim3.24$ show similar SFRs 
as those of  \oiii\ emitters at $z\sim2.23$ 
at a fixed stellar mass. 
The distribution of the \oiii\ candidate emitters at $z\sim2.23$ is consistent 
with the fit to the \oiii\ candidate emitters at $z\sim3.24$ statistically. 
While the normalization of the stellar mass--SFR relation is almost consistent, 
the distribution along the relation seems to be different.  
The \oiii\ emitters at $z\sim3.24$ show an offset towards the lower stellar mass range 
as seen in the top and right panels of Figure~\ref{fig:SFMS_all} 
(\citealt{suzuki15}; comparison between the \oiii\ emitters at $z\sim3.2$ and the \ha\ emitters at $z\sim2.2$).

%%%

\subsection{Stacking analysis}\label{subsec:stack}

In order to investigate the averaged properties of 
the \oiii\ emitters at $z\sim3.2$, 
we carry out the stacking analysis of the spectra 
by dividing the ten \oiii\ emitters into two stellar mass bins, 
i.e. $9.76 \le \mathrm{log(M_*/M_\odot)} \le 10.21$ and $9.07 \le \mathrm{log(M_*/M_\odot)} \le 9.23$.  
 
\begin{figure*}
\centering\includegraphics[width=0.65\textwidth]{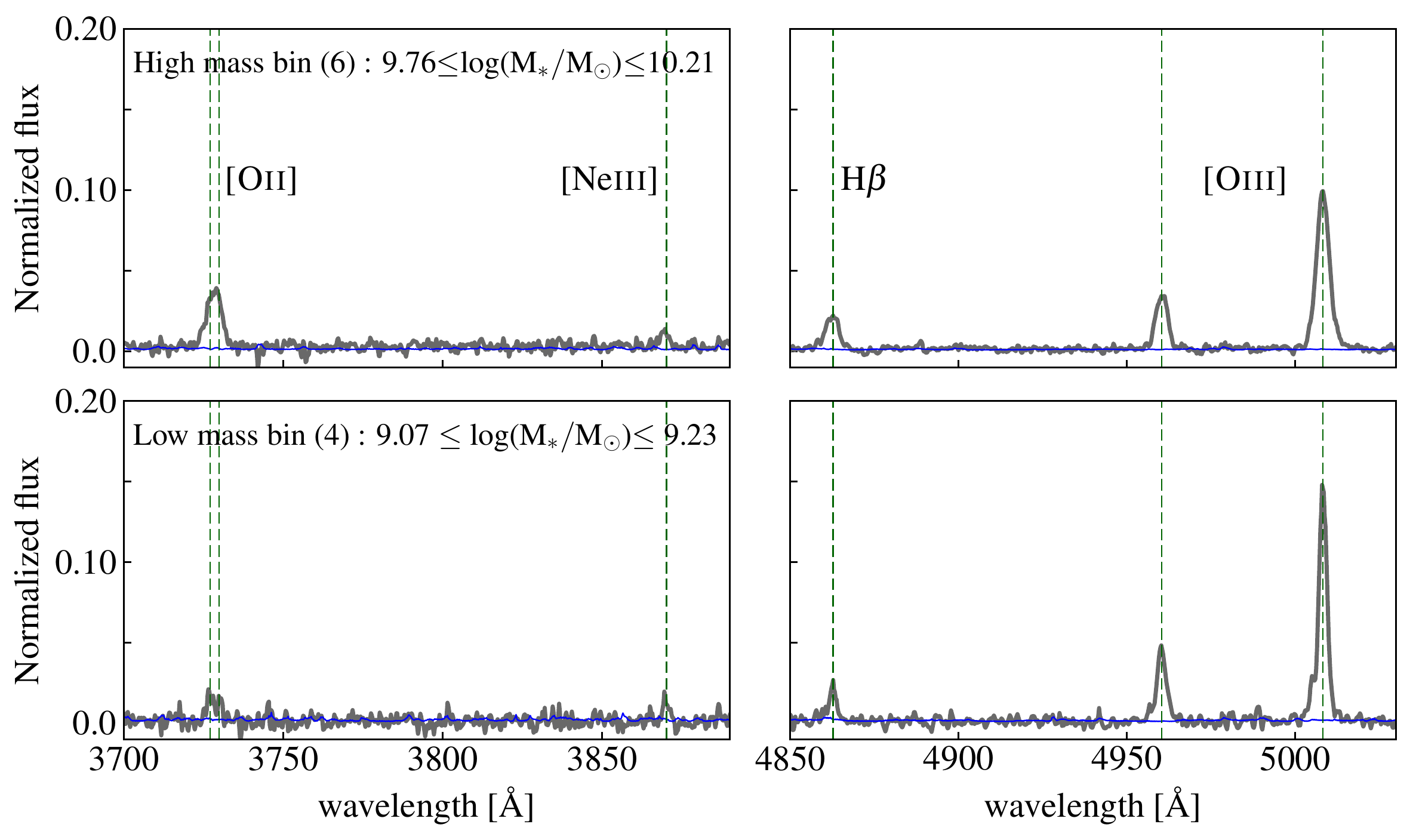}
\caption{
Stacked spectra of the \oiii\ emitters 
obtained by dividing the samples into two stellar mass bins 
of $9.76 \le \mathrm{log(M_*/M_\odot)} \le 10.21$ and $9.07 \le \mathrm{log(M_*/M_\odot)} \le 9.23$. 
The stacked spectra are shown by the gray curves. 
The blue curves represent the 1$\sigma$ sky noise. 
We show the \oii\ doublet and the \neiii\ in the left panel, 
and the \hb\ and the \oiii\ doublet in the right panel, respectively. 
}
\label{fig:stacked}
\end{figure*}

We transform the individual spectra to the rest-frame wavelength 
based on the derived redshifts, 
and normalize them by integrated \oiii$\lambda$5008 flux.  
The wavelength dispersion of the spectrum in $K$ and $H$-band is 2.1719~\AA/pix and 1.6289~\AA/pix, respectively. 
When converting them to the rest-frame spectra,  
we fix the wavelength interval to 0.25~\AA, and interpolate the spectra linearly. 
Noise spectra for the individual galaxies 
are also scaled by integrated \oiii$\lambda$5008 flux, 
and are similarly converted to the rest-frame wavelength. 
Then, the stacking of the individual spectra is carried out with the following equation: 

\begin{equation}
f_{\mathrm{stack}} = \sum_{i}^N \frac{f_i (\lambda)}{\sigma_{i} (\lambda)^2}/\sum_i^N \frac{1}{\sigma_i (\lambda)^2}, 
\end{equation} 

\noindent
where $f_i (\lambda)$ is a flux density of the individual spectra  
and $\sigma_i (\lambda)$ is a sky noise as a function of the wavelength \citep{shimakawa15}. 
The noise spectrum for the stacked spectrum is calculated 
by an error propagation from the individual noise spectra. 
The stacked spectra in the two stellar mass bins are shown in Figure \ref{fig:stacked}.

%% Figures of individual and stacked spectra

\section{ISM conditions of [{\sc OIii}] emitters among other samples at $z>3$}\label{sec:z3}

\subsection{Line ratios and its stellar mass-dependence at $z>3$}\label{subsec:ism}

\begin{figure*}
\centering\includegraphics[width=0.7\textwidth]{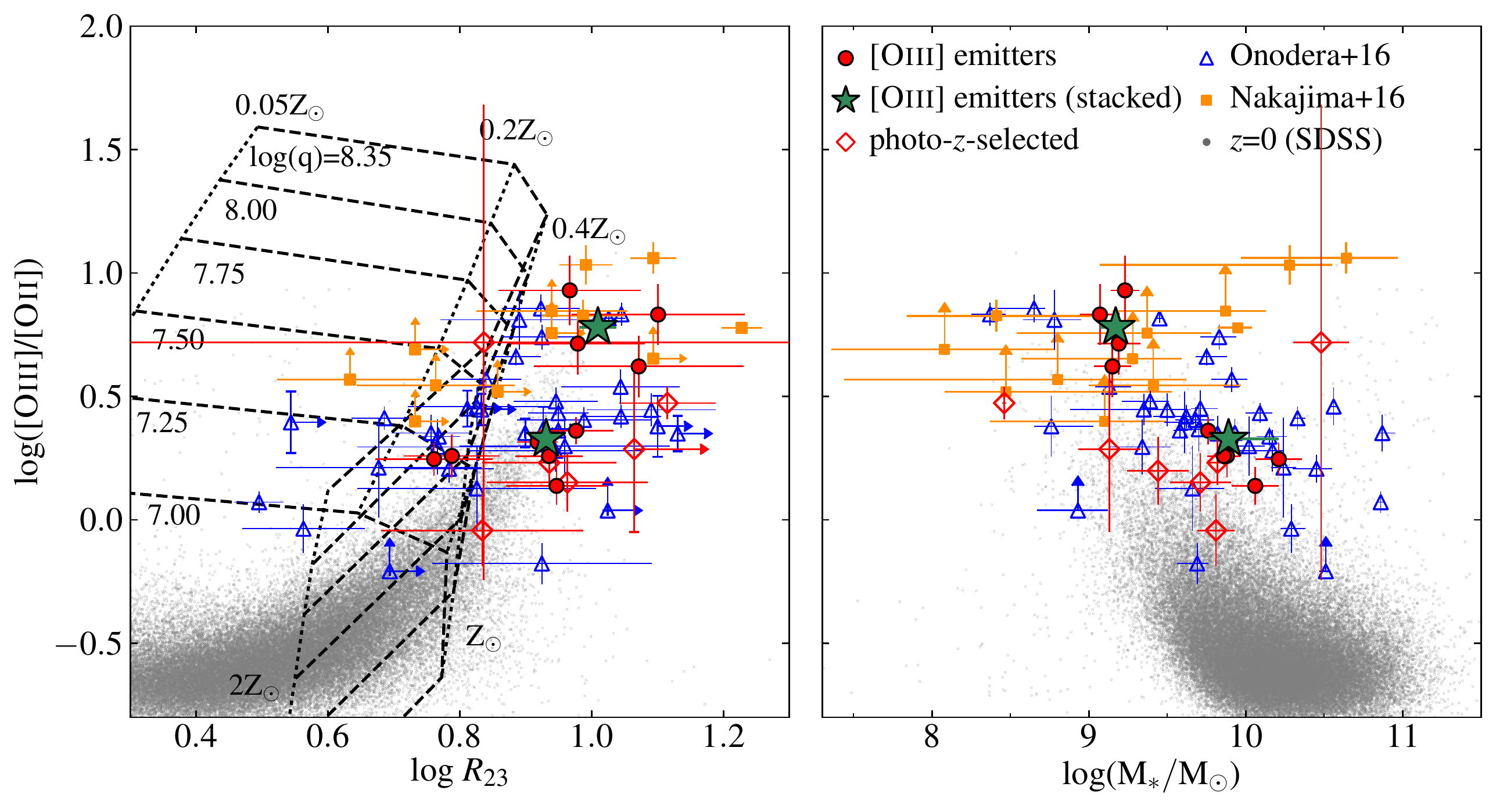}
\caption{
Relation between the $R_\mathrm{23}$-index and \oiii$\lambda\lambda$5008,4960/\oii\ ratio (left) and 
between the stellar mass and the \oiii$\lambda\lambda$5008,4960/\oii\ ratio  
 (right) 
of our sample at $z\sim3.2$, \oiii\ emitters and photo-$z$-selected sources. 
We also plot UV-selected star-forming galaxies at $z=3-3.7$
from \citet{onodera16}, LAEs at $z\sim3$ from \citet{nakajima16}, 
and 
local star-forming galaxies from SDSS \citep{sdssdr7,sdssdr8}.
In the left panel, 
the dashed and dotted lines represent the model prediction of the $R_{23}$-index and the \oiii/\oii\ ratio 
calculated using the photoionization code {\tt MAPPINGS V}. 
Star-forming galaxies at $z>3$ have different ISM conditions 
from those of local star-forming galaxies. 
Comparing among samples at $z>3$, 
massive \oiii\ emitters ($\mathrm{log(M_*/M_\odot)} \sim$ 9.8--10.2) 
seem to show similar line ratios as UV-selected galaxies,  
while less massive \oiii\ emitters ($\mathrm{log(M_*/M_\odot)} \sim 9.0$) are similar to LAEs. 
When \hb\ is detected with S/N $<3.0$, 
we replace it with the 3$\sigma$ flux limit.  
The source not detected with \hb\ is not shown in the left panel.
}
\label{fig:ism}
\end{figure*}

The left panel of Figure~\ref{fig:ism} shows  
the relation between two line ratios, namely, 
the $R_\mathrm{23}$-index 
((\oiii$\lambda \lambda$5008,4960 + \oii) / \hb)
and \oiii$\lambda\lambda$5008,4960/\oii\ ratio. 
While the $R_{\rm 23}$--index and \oiii/\oii\ ratio depend on 
both the gas metallicity and ionization parameter, 
the $R_{\rm 23}$ is more sensitive to the gas metallicity 
and \oiii/\oii\ is more sensitive to the ionization parameter \citep[e.g.][]{kewleydopita02,nakajima14}. 

We show our sample on the $R_{\rm 23}$--\oiii/\oii\ diagram 
together with star-forming galaxies at the same epoch from the literature, 
namely, UV-selected galaxies from \citet{onodera16} 
and LAEs from \citet{nakajima16}. 
The model predictions are also shown on the diagram. 
The theoretical line ratios in the {\sc Hii} regions are estimated using the photoionization code 
{\tt MAPPINGS V} \footnote{\url{https://miocene.anu.edu.au/mappings/}} ({\tt MAPPINGS}; \citealt{mappings}).  
In the {\tt MAPPIGNS}, 
we assume a {\sc Hii} region with a constant pressure of  
$P/k=10^{6.5} {\rm cm^{-3}\ K}$, 
where $k$ is the Boltzmann constant. 
The temperature of the {\sc Hii} region is set to be $\sim 10^4$\ K, and then 
the density becomes $\sim$ 300 ${\rm cm^{-3}}$, 
which corresponds to the typical electron density of 
star-forming galaxies at high redshifts 
\citep[e.g.][]{steidel14,shimakawa15,sanders16a,onodera16,strom16}.
We change the metallicity and ionization parameter independently as follows: 
$Z = 0.05,\ 0.2,\ 0.4,\ 1.0,\ 2.0\ Z_\odot$, 
and log($q\ [{\rm cm s^{-1}}]$) = 8.35, 8.00, 7.75, 7.50, 7.25, and 7.00.

In this paper, 
we use the ionization parameter defined as:  
 
 \begin{equation}
 q = \frac{Q_{\rm H^0}}{4 \pi R_{\rm s}^2 n_{\rm H}},
 \end{equation}
 
\noindent
where $Q_{\rm H^0}$ is the flux of the ionizing photons produced by the existing stars 
above the Lyman limit, 
$R_{\rm s}$ is the Str\"omgren radius, 
and $n_{\rm H}$ is the local density of hydrogen atoms 
(\citet{kewleydopita02}; 
and see also \citet{sanders16a} for detailed discussions about the definitions of the ionization parameter).

In the right panel of Figure~\ref{fig:ism}, 
we show the relation between the stellar mass and 
the \oiii$\lambda\lambda$5008,4960/\oii\ ratio 
of the same samples shown in the left panel 
in order to clarify the differences in the stellar mass distributions among the samples.

In Figure~\ref{fig:ism}, 
we also show local star-forming galaxies from SDSS Data Release 8 (DR8), 
whose physical quantities are provided by the MPA-JHU group\footnote{\url{http://wwwmpa.mpa-garching.mpg.de/SDSS/}} 
\citep{sdssdr7,sdssdr8}.
We clearly see that star-forming galaxies at $z>3$ show very 
different line ratios from those of local star-forming galaxies, 
in the sense that those of $z>3$ galaxies 
tend to have higher \oiii/\oii\ ratios at a fixed 
$R_\mathrm{23}$-index and stellar mass. 
This confirms the results already reported in the literature using the UV-selected galaxies 
that the ionization states of 
star-forming galaxies at $z>3$ are higher than those of star-forming galaxies at $z=0$ 
\citep[e.g.][]{holden16,onodera16,nakajima16}.

When we compare our sample to the sample of \citet{onodera16} 
in Figure~\ref{fig:ism}, 
there is no clear difference between the two samples. 
The \oiii\ emitters are not systematically biased towards 
higher $R_\mathrm{23}$-index or higher \oiii/\oii\ ratios 
with respect to the UV-continuum-selected star-forming galaxies at the same epoch.  
When comparing the LAEs at $z\sim3$ from \citet{nakajima16}, 
at a lower stellar mass regime of $\mathrm{log(M_*/M_\odot)} \sim 9.0$,  
the \oiii\ emitters are likely to be consistent with being 
the same population as LAEs. 
Our results suggest that the selection based on the \oiii\ emission line strength 
does not cause any significant bias in terms of the ISM conditions, 
and moreover, that we can pick up star-forming galaxies 
in a wide range of ISM conditions 
from ones with extreme conditions such as LAEs 
to ones with moderate conditions at $z>3$.

\subsection{Metallicity estimation with the empirical calibration method}\label{subsec:empirical}

We use the fully empirical relations   
calibrated using local star-forming galaxies from SDSS by \citet{curti16}.
They introduced the empirical relations between the gaseous metallicities 
and six line ratios, 
and in this study, 
we use four line ratios with \oiii, \hb, and \oii\ lines. 
Hereafter, 
we estimate gas metallicities only for 
 the sources with all of these emission lines being detected with S/N $\ge3$. 
Also, we remove the source with a large uncertainty of $\rm A_{FUV}$.  
Note that all of the removed sources are the photo-$z$-selected sources.

\begin{figure*}
\centering\includegraphics[width=1.0\textwidth]{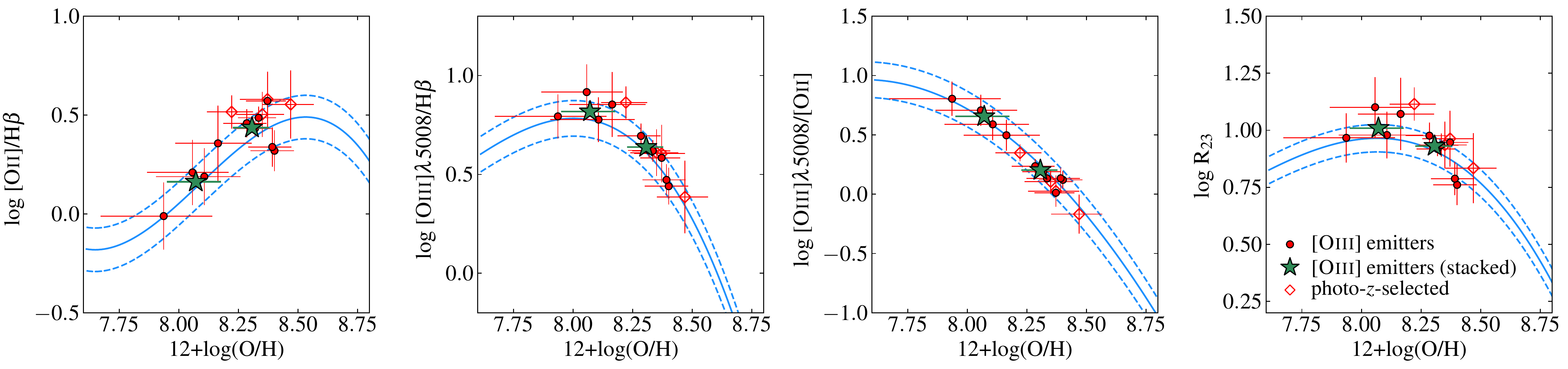}
\caption{
Relations between four line ratios and metallicities calibrated with the \citet{curti16} method.
The solid curve in each panel represents the relation derived in \citet{curti16}. 
The dashed curves represent the root-mean-square of their fit.   
The four line ratios of individual galaxies are well fitted by their empirical relations 
within 1$\sigma$ errors. 
}
\label{fig:lineratio_C16}
\end{figure*}

We fit the four line ratios simultaneously, 
and determine the best-fit metallicity that can minimize the $\chi^2$ value. 
Here the $\chi^2$ is defined as follows:

\begin{equation}
\chi ^{2} = \sum_{i=1}^N \frac{(\mathrm{log}\ R_{i,\mathrm{obs}} - \mathrm{log}\ R_{i,\mathrm{fit}})^2}{\sigma^2 _{i,\mathrm{obs}} + \sigma^2 _{i,\mathrm{int}}}, 
\end{equation}

\noindent
where 
log $R_{i, \mathrm{obs}}$ and log $R_{i,\mathrm{fit}}$ 
are the $i$-th line ratio obtained from the observed spectra 
and one obtained from the relation of \citet{curti16} 
at a given metallicity \citep{onodera16}. 
$\sigma_{i,\mathrm{obs}}$ is the error of each line ratio from the observed spectra, and 
$\sigma_{i,\mathrm{int}}$ is the intrinsic scatter of a line ratio at a given metallicity, respectively. 
We apply the root-mean-square estimated for each relation (Table~2 in \citet{curti16}) 
as the intrinsic scatter.   
%% error estimation 
In Figure~\ref{fig:lineratio_C16}, 
we show the relations between the metallicity, which is 
determined with two different calibration methods, 
and line ratios. 
Note that the four line ratios shown in Figure~\ref{fig:lineratio_C16} are not independent, 
and 
the 1$\sigma$ errors in the metallicities are 
determined from values of 12+log(O/H) with $\Delta \chi^2 = 3$ 
compared to the best fit solution.

We note that locally calibrated relations between line ratios and gas metallicity 
might not be applicable to star-forming galaxies at high redshifts 
because their typical ISM conditions seem to change from $z=0$ 
(\citealt{kewley13,nakajima14,steidel14,strom16,kashino17} and Figure~\ref{fig:ism}), 
while some previous studies have suggested that physical conditions of {\sc Hii} regions do not evolve with redshifts 
at a fixed metallicity  
\citep[e.g.][]{jones15,sanders16}.   
Nevertheless, 
since it is shown that the gas metallicities estimated with different line ratios 
show systematic offsets from one another \citep{kewleyellison08},  
we here use the locally calibrated empirical relations to estimate gas metallicities 
for a fair comparison with \citet{onodera16} in the next section.

\subsection{Mass--Metallicity relation at $z>3$}\label{subsec:mz}

\begin{figure*}
\centering\includegraphics[width=1.0\columnwidth]{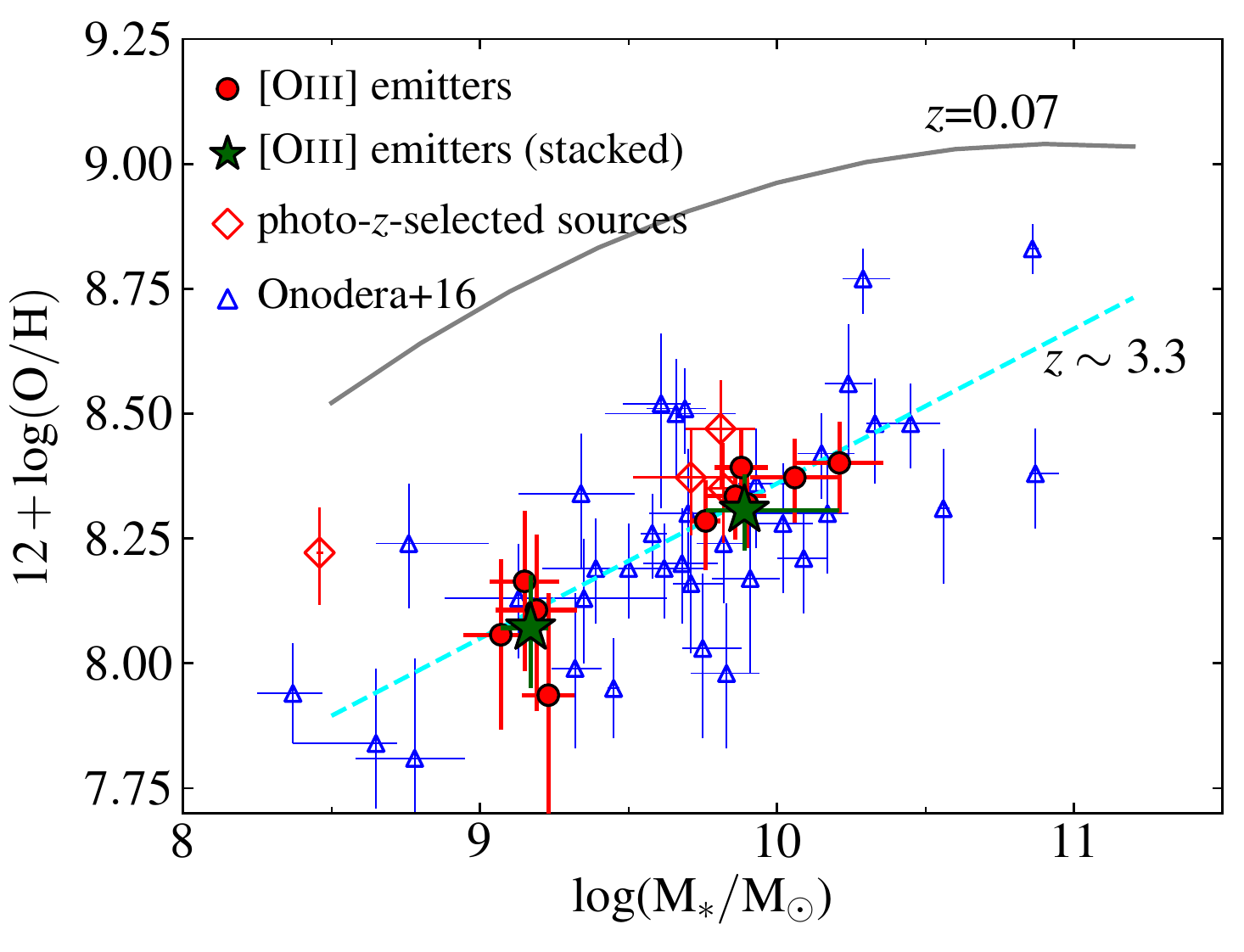}
\caption{
Relation between stellar mass and gas metallicity for our sample at $z\sim3.2$   
and the UV-selected galaxies at $z\sim$ 3--3.7 from \citet{onodera16}.  
The solid curve represents the mass--metallicity relation at $z=$ 0.07 \citep{maiolino08}. 
The dashed curve represents the best-fitted mass--metallicity relation at $z\sim3.3$ from \citet{onodera16}. 
Our targets are well below the mass--metallicity relation of the local star-forming galaxies. 
Comparing with the UV-selected galaxies at the same epoch, 
 there is no clear difference of gas metallicities at a fixed stellar mass between the two samples. 
 Our \oiii\ emitters well follow the best-fitted relation by \citet{onodera16}. 
}
\label{fig:MZ_z3}
\end{figure*}

In Figure \ref{fig:MZ_z3}, 
we show the relation between stellar mass and gas metallicity for our sample.  
As already shown in a number of previous studies, 
stellar mass and metallicity of our galaxies at $z\sim3.2$   
show a correlation such that more massive galaxies have higher metallicities 
(e.g. \citealt{tremonti04, erb06-644, maiolino08, stott13, zahid13, zahid14, steidel14, troncoso14, sanders15}). 
UV-selected galaxies at the same epoch from the \citet{onodera16} are also shown.  
We find no clear difference of gas metallicities between the \oiii\ emitters and the UV-selected galaxies 
at a fixed stellar mass.  

As also suggested in Figure~\ref{fig:ism}, 
\oiii\ emitters are not biased towards a particular population 
with respect to their ISM conditions and metal contents as compared to the UV-continuum-selected galaxies 
at least in the stellar mass range covered by our observation, i.e. $\mathrm{log(M_*/M_\odot)} \sim $ 9.0--10.2. 
%%% 
It is expected that the effect of dust extinction is not significant in our stellar mass range, and 
therefore, there is no difference between the \oiii-selected and the UV-selected galaxies. 
If the \oiii-selected galaxies can trace more massive and dustier star-forming galaxies,  
the difference might appear in more massive stellar mass range, 
and a larger sample of the \oiii\ emitters and their follow-up observations are required. 
%%%

\section{Comparison with star-forming galaxies at $z\sim2$}\label{sec:comparewithz2}

\subsection{Metallicity calibration based on photoionization modelling}\label{subsec:kk04}

We apply the calibration method, which is introduced by \citet[][KK04]{KK04},  
as well as the empirical calibration method by \citet{curti16} as described in Section \ref{subsec:empirical} 
in order to compare our sample with previous studies at $z\sim2$ in the following sections.

\citetalias{KK04} used strong emission lines and determined relations between line ratios, gas metallicities and ionization parameters 
based on the photoionization model, {\tt MAPPINGS}. 
In this method,  
the gas metallicity and ionization parameter are determined simultaneously 
using the two line ratios of the $R_{\mathrm{23}}$-index and \oiii/\oii. 

We estimate the gas metallicity and ionization parameter by following \citetalias{KK04}. 
The relation between ionization parameter $\mathrm{log(q)}$ 
and \oiii$\lambda\lambda$5008,4960/\oii\ ratio is given by

\begin{eqnarray}
 %\begin{split}
 & \mathrm{log} & (q)   =  \{32.81 - 1.153 y^2   \label{eq:logq} \\
 & + & [12+\mathrm{log(O/H)}](-3.396 - 0.025y + 0.1444 y^2)    \nonumber  \\
 & \times & \{4.603 - 0.3119y -0.163y^2 \nonumber  \\
 & + & [12+\mathrm{log(O/H)}](-0.48 + 0.0271y + 0.02037y^2)\}^{-1}, \nonumber
 %\end{split}
 \end{eqnarray}

\noindent 
where $y = \mathrm{log([\textsc{Oiii}]\lambda\lambda5008,4960/[\textsc{Oii}])}$. 
The relation between gas metallicity 12+log(O/H) and the $R_{\mathrm{23}}$-index 
is separated into the two equations according to gas metallicity. 
At the lower metallicity branch of ${\rm 12+log(O/H)} < 8.4$, 

\begin{eqnarray}
 12 +  \mathrm{log(O/H)_{lower}} & =  & 9.40 + 4.65x - 3.17x^2 \label{eq:lowerZ} \\
 & - &  \mathrm{log}(q)(0.272 + 0.547x - 0.513x^2), \nonumber 
\end{eqnarray}

\noindent 
and at the upper metallicity branch of ${\rm 12+log(O/H)} \ge 8.4$, 

\begin{eqnarray}
12 & + &  \mathrm{log(O/H)_{upper}} =  9.72 - 0.777x - 0.951x^2  \nonumber  \\
& -  & 0.072x^3 - 0.811x^4  
-  \mathrm{log}(q) (0.0737 - 0.0713x  \nonumber \\
& - & 0.141x^2 + 0.0373x^3 - 0.058x^4),   \label{eq:upperZ} 
 \end{eqnarray}

\noindent 
where $x= \mathrm{log} R_{\rm 23}$. 
Consistent metallicity and ionization parameter are determined in an iterative manner 
using Eq.(\ref{eq:logq}) and Eq.(\ref{eq:lowerZ}) or Eq.(\ref{eq:upperZ}) 
according to the value of 12+log(O/H) \citepalias{KK04}.

We compare gas metallicities obtained by the \citetalias{KK04} method 
with those obtained in Section~\ref{subsec:empirical}. 
When we see the upper metallicity branch, 
the gas metallicities based on the photoionization models 
are systematically higher ($\sim 0.25\ {\rm dex}$)  
than those from the empirical relations.  
As for the solutions at the lower metallicity branch, 
there is no systematic offset with respect to the results from the 
empirical relations 
but they seem to show a negative trend with respect to the stellar mass 
(Appendix~\ref{appendix:twometal}).

In order to determine the metallicity branch at a given $R_{\rm 23}$-index, 
an additional line ratio, such as \nii/\oii, is required \citepalias{KK04}. 
Since we cannot observe \nii$\lambda$6585 lines for $z>3$ galaxies from the ground, 
it is difficult to determine the metallicity branch for each object in our sample.  
In the following sections, 
we only show the gas metallicities at the upper branch 
for clarity.

We note that 
\citet{steidel14} suggested the possibility that 
metallicity calibration methods using the $R_\mathrm{23}$-index 
do not work well in the metallicity range of $\mathrm{12+log(O/H)}=$ 8.0--8.7. 
However, here we use the \citetalias{KK04} method due to the limited available emission lines 
of our sample 
and also for a fair comparison with previous studies at $z\sim2$.  
\citet{kewleyellison08} showed that the gas metallicities with the calibration methods using different line ratios 
show systematic offsets from one another. 
Therefore, we attempt to compare gas metallicities estimated with the same calibration method.

\subsection{Comparison of the ionization parameter and gas metallicity}\label{subsec:qzcomparison}

In Figure \ref{fig:kk04} (a), 
we show gas metallicities and ionization parameters 
of our sample estimated in Section \ref{subsec:kk04}. 
Here we show the two solutions at the upper and lower metallicity branch 
while some sources have the same solution at the two branches 
indicating that they lie at the cross-over metallicity.  
We also show the results of LBGs  
and LAEs at $z\sim$ 2--3 from \citet{nakajima14}, 
who estimated gas metallicities and ionization parameters with the \citetalias{KK04} method. 
Comparing our sample with LBGs and LAEs of \citet{nakajima14} on this diagram, 
our sample at $z\sim3.2$ shows similar gas metallicities and ionization parameters 
as those of the LBGs at $z\sim$ 2--3. 

In Figure \ref{fig:ism}, 
we find that star-forming galaxies at $z>3$ clearly show different line ratios 
from those of the local star-forming galaxies, 
indicating that they are likely to have higher ionization parameters 
at a fixed metallicity or stellar mass. 
Figure \ref{fig:kk04} (a) 
indicates that the redshift evolution of ISM conditions is unlikely to be strong 
between $z\sim3.2$ and $z\sim2$. 
The sample of LBGs of \citet{nakajima14} 
covers a wider stellar mass range than that of our sample,  
$\mathrm{log(M_*/M_\odot) = 8.0-10.8}$.  
We also note that their LBG sample 
includes galaxies at $z\sim3$ from AMAZE \citep{maiolino08}, 
and this might contribute to similar ionization parameters and gas metallicities 
between the two samples.

\subsection{Comparison of mass--metallicity relation}\label{subsec:comparisonz2}

In Figure~\ref{fig:kk04} (b), 
we show the relation between stellar mass and gas metallicity again, 
but gas metallicities are estimated with the \citetalias{KK04} method 
for a fair comparison with previous studies about star-forming galaxies at $z\sim2$ 
\citep{zahid13,cullen14,steidel14,sanders15}.

We introduce some previous studies at $z\sim2$.  
\citet{cullen14} investigated ISM conditions of star-forming galaxies at $z\sim2.2$ 
selected from the 3D-{\it HST} grism survey data. 
Their sample is basically selected by their strong \oiii\ emission lines. 
They stacked their samples into six stellar mass bins 
and measured the fluxes of the \oii, \hb, and \oiii\ lines. 
We here directly estimate gas metallicities of their sample with the \citetalias{KK04} method. 
We show the solutions 
at the upper metallicity branch in Figure~\ref{fig:kk04} (b). 

We also show the results from \citet{steidel14} and \citet{sanders15}, 
who calibrated gas metallicities using the \nii/\ha\ lines ratios (N2) by \citet[][PP04]{PP04}. 
We converted their gas metallicities using the formula given by \citet{kewleyellison08} 
so that gas metallicities correspond to those estimated using the \citetalias{KK04} method. 
We show one more previous study, \citet{zahid13}. 
They obtained the mass--metallicity relation at $z\sim2.2$ with the \citetalias{KK04} method 
by converting the mass--metallicity relation obtained by \citet{erb06-644} with the N2 \citepalias{PP04} method 
with the formula by \citet{kewleyellison08}.

The thick dash-dotted line in Figure~\ref{fig:kk04} (b) 
shows the best-fitted mass--metallicity relation 
derived using the solutions at the upper metallicity branch 
of our sample at $z\sim3.2$. 
We compare this best-fitted relation at $z\sim3.2$ with that 
estimated for \citet{cullen14} sample. 
The slopes and intercepts of the best-fitted lines for the two samples 
are consistent with each other within errors, 
indicating that 
the gas metallicities of our sample at $z\sim3.2$ 
are similar those of star-forming galaxies at $z\sim2.2$ 
at a fixed stellar mass. 
This is also the case when comparing the solutions at the lower metallicity branch 
(Appendix~\ref{appendix:twometal}).

On the other hand, 
comparing with other previous studies, 
which estimated gas metallicity with N2 \citepalias{PP04} method originally, 
they tend to have higher metallicities with respect to our sample and the sample of \citet{cullen14}. 
It is suspected that there is still a systematic difference due to using different calibration methods even after the correction. 
The correction factors for local star-forming galaxies introduced in \citet{kewleyellison08} 
might not be applicable for star-forming galaxies at $z>2$ 
due to their different physical conditions. 
Therefore, comparing our targets with the samples whose metallicities are originally calibrated by the N2 \citepalias{PP04} method 
 might not be fair. 
We conclude that our sample at $z\sim3.2$ has similar ISM conditions and mass--metallicity relation 
as star-forming galaxies at $z\sim2$ under the same calibration method.

\subsection{ISM conditions and star-forming activity between $z\sim3.2$ and $z\sim2$}\label{sec:discussion}

\begin{figure*}[tbp]
\begin{minipage}[cbt]{0.5\textwidth}
\centering\includegraphics[width=0.9\columnwidth]{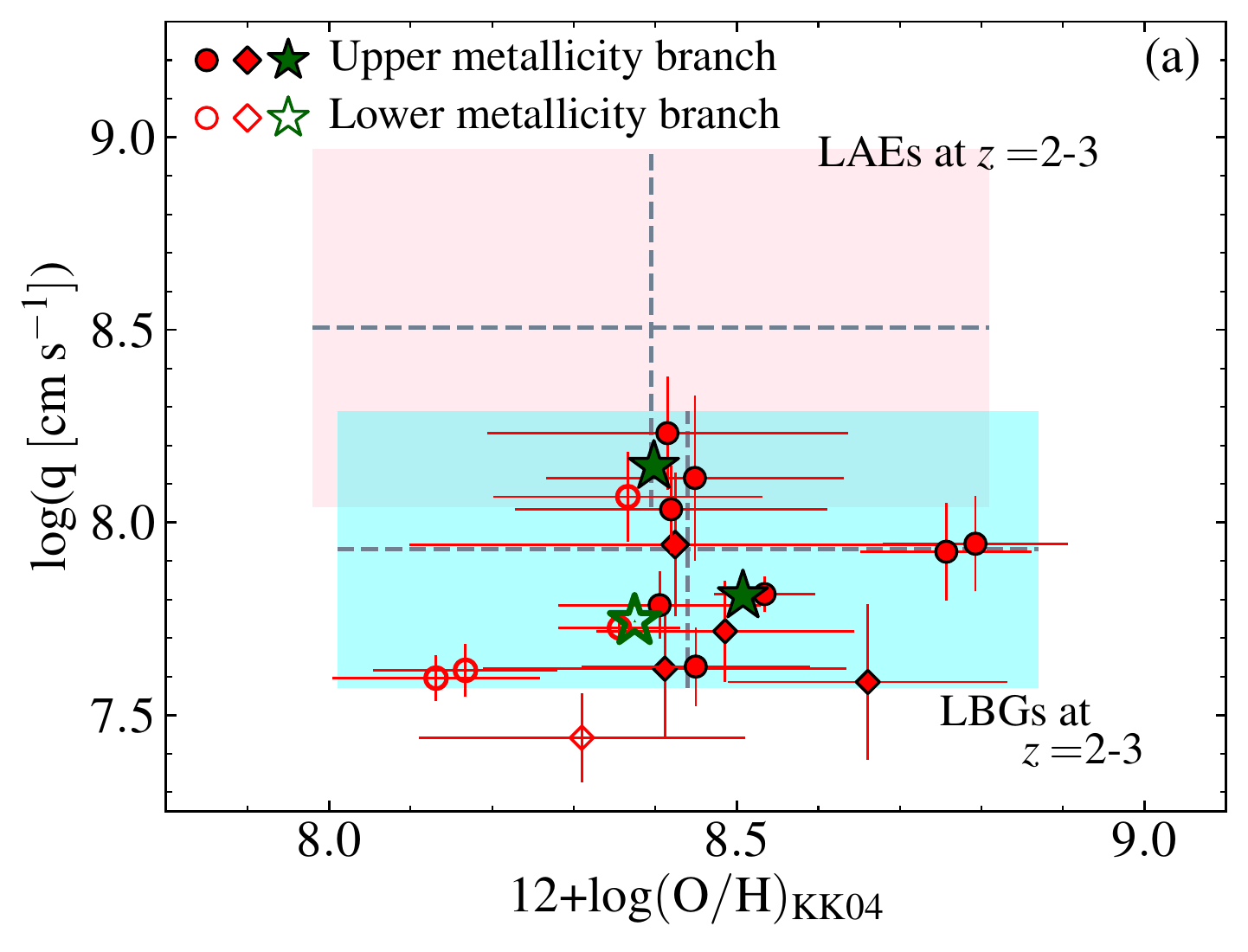}
\end{minipage}
\begin{minipage}[cbt]{0.5\textwidth}
\centering\includegraphics[width=0.9\columnwidth]{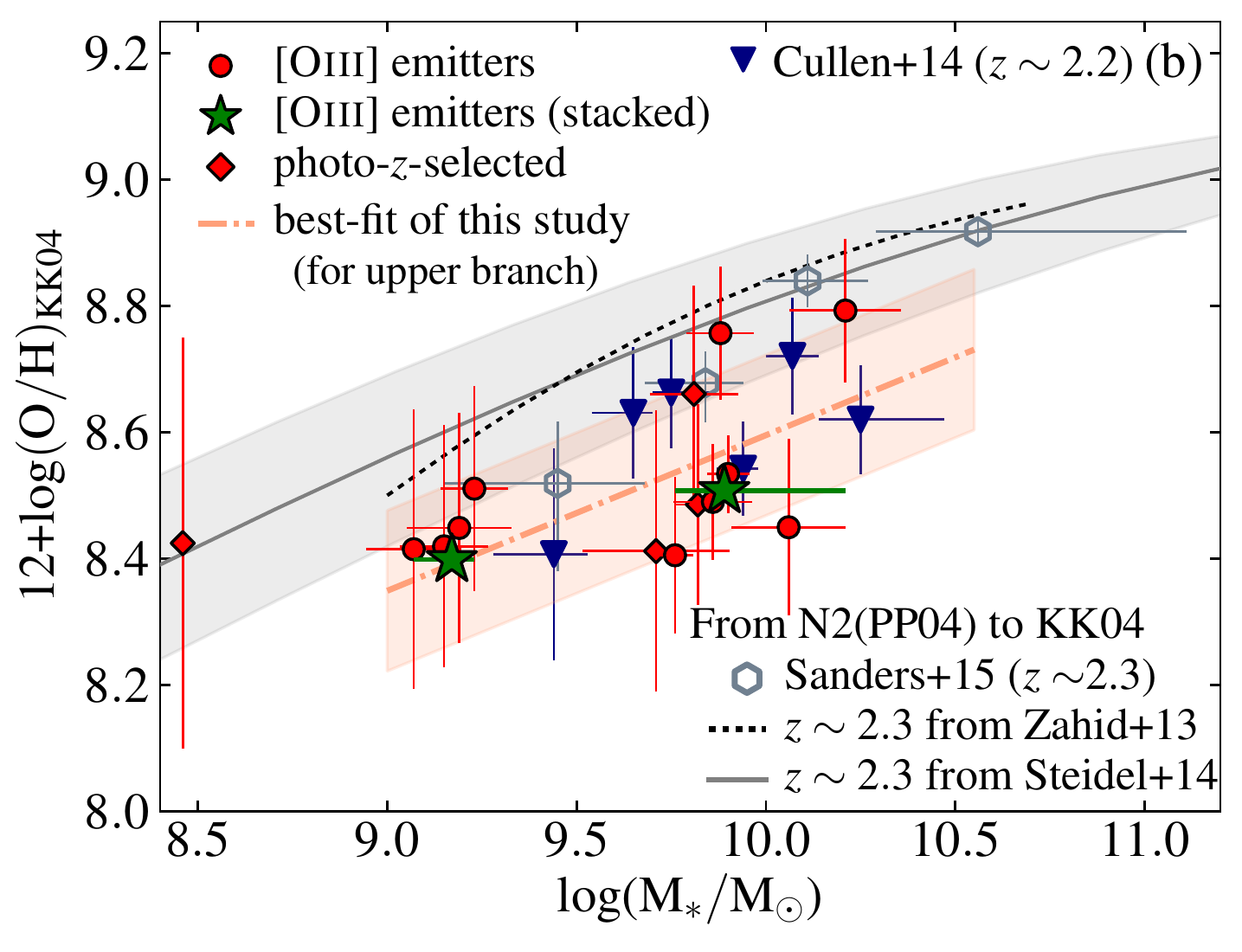}
\end{minipage}
\caption{
(a) Gas metallicity and ionization parameter of our sample at $z\sim3.2$ estimated using the \citetalias{KK04} method 
(circles: \oiii\ emitters and diamonds: photo-$z$-selected sources).  
The two solutions are shown with the filled (upper metallicity branch) and open (lower metallicity branch) symbols. 
The stacking results of the \oiii\ emitters are shown with the star symbols. 
The blue and red shaded region shows the results of the LBGs and LAEs at $z\sim$ 2--3, respectively, 
from \citet{nakajima14}. 
The same method is applied here. 
Our targets at $z>3$ seem to cover a similar range of gas metallicity and ionization parameter 
as that of the LBGs at $z\sim$ 2--3.  \\
(b) Relation between stellar mass and gas metallicity for our sample at $z\sim3.2$. 
We estimate gas metallicities with the \citetalias{KK04} method here, and 
the solutions at the upper metallicity branch are shown.  
The dash-dotted line shows the best-fitted line derived 
for our sample.
We compare our sample at $z\sim3.2$ with 
previous studies about star-forming galaxies at $z\sim2$.
Except for the sample of \citet{cullen14}, 
gas metallicities are originally calibrated with N2 \citepalias{PP04} method  
and then are converted using a formula by \citet{kewleyellison08} 
\citep{zahid13,steidel14,sanders15}. 
The red and gray shaded region corresponds to $\pm 1\sigma$ errors of 
the best-fitted relation of our sample using results at upper metallicity branch 
and \citet{steidel14}, respectively. 
The mass--metallicity relation of our sample is consistent with that of \citet{cullen14} 
within 1$\sigma$ error. 
}
\label{fig:kk04}
\end{figure*}

In Figure~\ref{fig:SFMS_all}, 
we show that the normalization of star-forming main sequence 
seems to be similar at 
$z\sim3.2$ and $z\sim2$.  
Also, as shown in Figure~\ref{fig:kk04}, 
it is suggested that the ISM conditions and the mass--metallicity relation 
do not seem to evolve between the two epochs. 
These results suggest that the properties of star-forming galaxies 
at $z\sim$ 2.0--3.2 (the difference of cosmic age of $\sim1.3$ Gyr) are primarily 
determined by their stellar masses rather than cosmic epoch 
since galaxies are very young and their ages are getting closer to the age of the Universe ($\sim$ a few Gyr).

As discussed in \citet{suzuki15} 
and as suggested distributions between the \oiii\ emitters at $z\sim3.2$ and $z\sim2.2$ 
along the main sequence  (Figure~\ref{fig:SFMS_all}), 
the individual galaxies should experience significant growth in stellar masses. 
These results probably reflect that galaxies are in the vigorous formation phase at this epoch, 
and such a significant growth must be supported by ample gas accretion 
from the outside throughout these early epochs \citep[e.g.][]{keres05,keres09,dekel09.nature,bouche10}.

\citet{onodera16} 
also showed that 
the gas metallicity difference between their sample at $z>3$ 
and \citet{cullen14} sample at $z\sim2.2$ is relatively small at a fixed stellar mass.  
They found that a simple gas regulator model with mildly evolving 
star formation efficiency \citep{lilly13}
could well predict the observational trend of the 
redshift evolution of the mass--metallicity relation.

By obtaining the gas mass fractions for our sample and combining them with gas metallicities and stellar masses, 
it will become possible to give constraints on the inflow and outflow rates by combining with gas metallicities and stellar masses 
\citep[e.g.][]{troncoso14,yabe15PASJ,seko16}. 
Dust continuum or CO line observations with ALMA will enable us to directly measure the molecular gas mass of individual galaxies at $z>3$.

\section{Summary}\label{sec:summary}

In this paper, we present the results from 
NIR spectroscopic follow-up of star-forming galaxies at $z\sim3.2$. 
Our primary targets are the NB-selected \oiii\ emission line galaxies 
obtained by HiZELS in the COSMOS field \citep{sobral13,khostovan15}. 
We obtain $H$ and $K$ band spectra 
of all ten \oiii\ emitters and seven photo-$z$-selected galaxies (our secondary targets). 
Our results demonstrate the high efficiency of follow-up observations of NB-selected galaxies 
with all candidates being confirmed as \oiii\ emitters. 
By exploiting our deep NIR spectra, 
we find that: 

\begin{enumerate} 
\item In comparison with local galaxies, 
our sample shows different ISM conditions, 
such as  
higher $R_\mathrm{23}$-index and higher \oiii/\oii\ ratio,  
and lower gas metallicity at a fixed stellar mass, 
consistent with many previous studies 
\citep[e.g.][]{troncoso14, nakajima14, steidel14, onodera16}.

\item We compare our spectroscopically confirmed galaxies at $z\sim3.0-3.5$ 
with other galaxy populations at similar redshifts \citep{onodera16,nakajima16}
on the $R_\mathrm{23}$-index -- \oiii/\oii\ ratio diagram 
and the stellar mass--\oiii/\oii\ ratio diagram. 
The \oiii\ emitters show broadly similar line ratios as UV-selected galaxies. 
Moreover, 
the line ratios of less massive \oiii\ emitters ($\mathrm{log(M_*/M_\odot)} \sim 9.0$) 
are consistent with those of LAEs. 
The \oiii-selection seems to cause no significant bias 
in terms of the ISM conditions, 
and the \oiii-selected galaxies can cover a wide range of 
stellar masses and ISM conditions of star-forming galaxies at $z>3$.  
The mass--metallicity relation of our sample is consistent with that of \citet{onodera16}.

 \item We also compare our sample at $z\sim3.2$ 
 with star-forming galaxies at $z\sim2$ from literature 
 \citep{zahid13,cullen14,nakajima14,steidel14,sanders15}. 
  Our sample shows similar ionization parameters, gas metallicities, and mass--metallicity relation 
  as those obtained by \citet{nakajima16} and \citet{cullen14} using the same calibration method. 
This suggests that the ISM conditions of star-forming galaxies do not strongly evolve at a fixed stellar mass 
between $z\sim3.2$ and $z\sim2.2$. 
Considering that the \oiii\ emitters at $z\sim3.2$ have similar SFRs as those at $z\sim2.2$ 
at a fixed stellar mass, 
our results support the idea that the evolutionary stages of star-forming galaxies, 
such as SFRs and ISM conditions, 
at $z\gtrsim2$ are primarily determined by their stellar masses rather than redshift.

 \end{enumerate}

Since our current spectroscopic sample is very small, 
it is necessary to carry out more observations on a larger sample 
in order to statistically reveal the evolution of ISM conditions 
and star-forming activities from $z>3$ to $z\sim2$. 
The low contamination of the NB-selected emitters 
will lead to high efficient follow-up observations 
making it ideal for such studies.

\acknowledgments

%% KAKENHI 
%% time  
We thank the anonymous referee for providing constructive comments. 
The spectroscopic data presented herein were obtained 
at the W.M. Keck Observatory, which is operated as a scientific partnership 
among the California Institute of Technology, the University of California and the National Aeronautics 
and Space Administration. 
The Observatory was made possible by the generous financial support of the W.M. Keck Foundation. 
Observations (S16A-058) were carried out within the framework of Subaru-Keck 
time exchange program, where the travel expense
was supported by the Subaru Telescope, which is operated by
the National Astronomical Observatory of Japan (NAOJ). 
The authors wish to recognize and acknowledge the very significant cultural role 
and reverence that the summit of Mauna Kea has always had within the indigenous Hawaiian community.  
We are most fortunate to have the opportunity to conduct observations from this mountain. 
We thank Carlos Alvarez and the rest of the Keck telescope staff  
for their help in the observation. 
Data analyses were in part carried out on the open use data analysis computer system at the Astronomy Data Center, ADC, of NAOJ. 
TLS acknowledges support from JSPS KAKENHI Grant Number JP16J07112. 
MO acknowledges support from JSPS KAKENHI Grant Number JP17K14257. 
IS acknowledges support from STFC (ST/P000541/1), 
the ERC Advanced Grant DUSTYGAL (321334) and 
a Royal Society/Wolfson Merit Award. 

Funding for SDSS-III has been provided by the Alfred P. Sloan Foundation, the Participating Institutions, the National Science Foundation, and the U.S. Department of Energy Office of Science. The SDSS-III web site is \url{http://www.sdss3.org/}.

SDSS-III is managed by the Astrophysical Research Consortium for the Participating Institutions of the SDSS-III Collaboration including the University of Arizona, the Brazilian Participation Group, Brookhaven National Laboratory, Carnegie Mellon University, University of Florida, the French Participation Group, the German Participation Group, Harvard University, the Instituto de Astrofisica de Canarias, the Michigan State/Notre Dame/JINA Participation Group, Johns Hopkins University, Lawrence Berkeley National Laboratory, Max Planck Institute for Astrophysics, Max Planck Institute for Extraterrestrial Physics, New Mexico State University, New York University, Ohio State University, Pennsylvania State University, University of Portsmouth, Princeton University, the Spanish Participation Group, University of Tokyo, University of Utah, Vanderbilt University, University of Virginia, University of Washington, and Yale University.

\vspace{5mm}
\facilities{Keck:I (MOSFIRE)}

\software{IRAF,  
         MosfireDRP, 
         EAZY \citep{brammer08}, 
         FAST \citep{kriek09b}, 
         MAPPINGS \citep{mappings}, 
         TOPCAT \citep{topcat}
         }




\appendix

\section{$H$ and $K$-band spectra}\label{sec:spectra}

\begin{figure*}
\centering\includegraphics[width=1.0\textwidth]{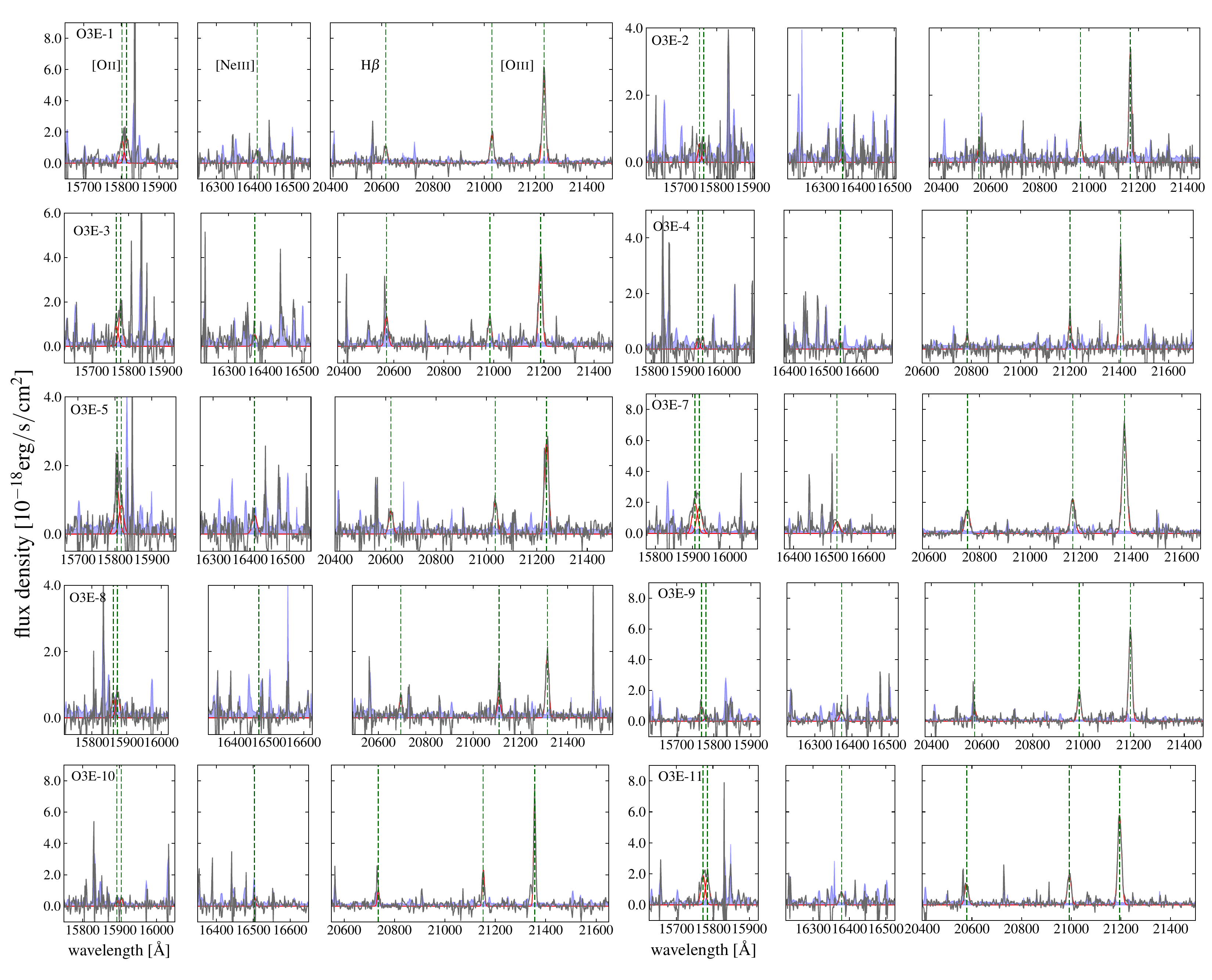}
\caption{
The $H$ and $K$-band spectra and the emission line fitting results of the ten \oiii\ emitters. 
The reduced spectra are shown with the gray curves. 
The blue shaded regions represent the 1$\sigma$ sky noise. 
The emission line fitting result with a Gaussian component is shown with the red curves for each source. 
Three panels show the emission lines, \oii$\lambda$3727, 
\oii$\lambda$3730 (left panel), [Ne{\sc iii}]$\lambda$3870 (middle panel), 
and \hb, \oiii$\lambda$4960, and \oiii$\lambda$5008 (right panel), respectively. 
We can see that the \oiii\ doublet, \hb, and \oii\ doublet lines are clearly detected with high signal-to-noise ratios 
for most of the \oiii\ emitters.  
}
\label{fig:HKspectra}
\end{figure*}

\begin{figure*}
\centering\includegraphics[width=1.0\textwidth]{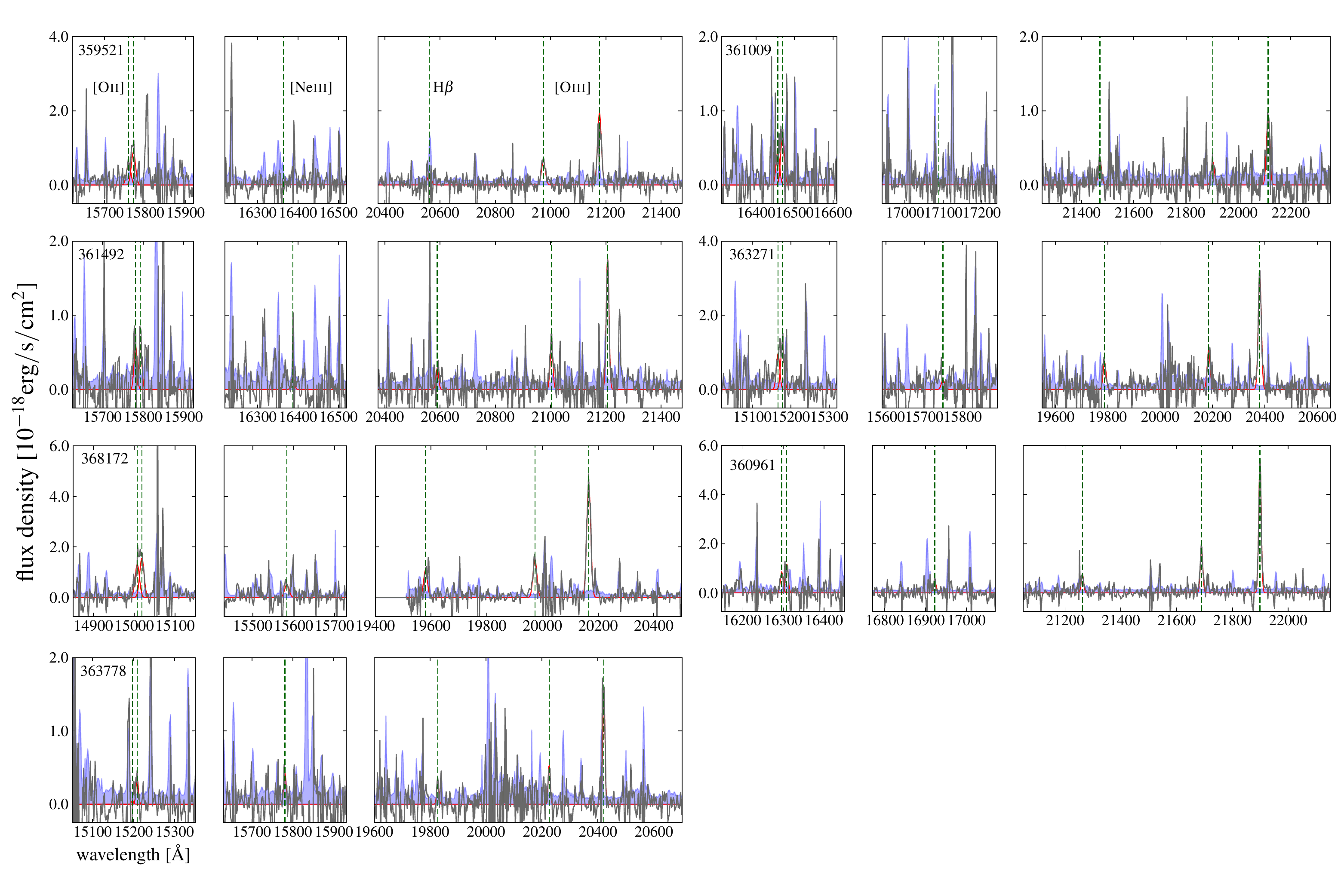}
\caption{
The $H$ and $K$-band spectra and the emission line fitting results 
of the seven photo-$z$-selected galaxies at $z_\mathrm{spec}=3.00-3.45$. 
he reduced spectra are shown with the gray curves. 
The blue shaded regions represent the 1$\sigma$ sky noise. 
The emission line fitting result with a Gaussian component is shown with the red curves for each source. 
Three panels show the emission lines, \oii$\lambda$3727, 
\oii$\lambda$3730 (left panel), [Ne{\sc iii}]$\lambda$3870 (middle panel), 
and \hb, \oiii$\lambda$4960, and \oiii$\lambda$5008 (right panel), respectively. 
Comparing the \oiii\ emitters (Figure~\ref{fig:HKspectra}), 
the \oiii\ fluxes of these galaxies are weaker.  
}
\label{fig:HKspectra_photoz}
\end{figure*}

In Figure~\ref{fig:HKspectra} and \ref{fig:HKspectra_photoz}, 
we show the $H$ and $K$-band spectra of the individual sources, 
namely the \oiii\ emitters ($z_\mathrm{spec}=$ 3.23--3.27) 
and the photo-$z$-selected galaxies ($z_\mathrm{spec}=$ 3.03--3.42). 
The emission line fitting results with a Gaussian component is shown in the red curves, 
and the results of the emission line fit are summarized in Table~\ref{table:spec-summary}. 
In Table~\ref{table:quantity-summary}, 
we summarize the estimated physical quantities, such as stellar masses, dust extinctions, 
$\mathrm{SFR_{UV}}$, 
and correction factors for stellar absorption for \hb.

%% Table of summary of redshifts and line fluxes
%% 
\begin{table*}
\begin{center}
\caption{
Summary of the emission line properties of the confirmed \oiii\ emitters and photo-$z$-selected sources with Keck/MOSFIRE.}
\label{table:spec-summary}
\begin{tabular}{ccccccccccc} \hline {\smallskip}
ID & $\mathrm{ID_{15}}$ & $\mathrm{ID_{S13}}$ & R.A. & Dec. & $z_{\rm spec}$ & FWHM & $F_\mathrm{[OIII]\lambda 5008}$ & $F_{\rm H\beta}$ & $F_{\rm [OII]}$ & $F_{\rm [NeIII]}$ \\
(1) & (2) & (3)  & (J2000) & (J2000) & & [$\rm km\ s^{-1}$] & (4) & (5) & (6) & (7)  \\ \hline
O3E-1 & 269781 & 7612 & 149.9485 & 1.6946 & 3.240 & 227 $\pm$ 6 & 9.38 $\pm$ 0.17 & 1.85 $\pm$ 0.16 & 4.0 $\pm$ 0.4 & 1.06 $\pm$ 0.17 \\ 
O3E-2 & 269719 & 7612 & 149.9777 & 1.6951 & 3.227 & 184 $\pm$ 9 & 4.63 $\pm$ 0.19 & 0.61 $\pm$ 0.20 & 1.08 $\pm$ 0.25 & $<$ 0.87 \\  
O3E-3 & 269241 & 7614 & 149.9751 & 1.6938 & 3.230 & 244 $\pm$ 11 & 7.19 $\pm$ 0.27 & 2.45 $\pm$ 0.45 & 3.5 $\pm$ 0.4 & 0.75 $\pm$ 0.19 \\ 
O3E-4 & 264007 & 7625 & 149.9418 & 1.6861 & 3.274 & 163 $\pm$ 5 & 4.43 $\pm$ 0.11 & 0.70 $\pm$ 0.08 & 0.76 $\pm$ 0.16 & $<$ 0.38 \\ 
O3E-5 & 260873 & 7632 & 149.9557 & 1.6804 & 3.241 & 285 $\pm$ 10 & 5.78 $\pm$ 0.22 & 1.42 $\pm$ 0.14 & 3.6 $\pm$ 0.4 & 0.91 $\pm$ 0.21 \\ 
O3E-7 & 293950 & 7569 & 149.9887 & 1.7333 & 3.268 & 309 $\pm$ 5 & 15.51 $\pm$ 0.21 & 3.51 $\pm$ 0.17 & 6.4 $\pm$ 0.4 & 1.41 $\pm$ 0.17 \\ 
O3E-8 & 293774 & 7569 & 149.9680 & 1.7332 & 3.256 & 232 $\pm$ 13 & 3.42 $\pm$ 0.19 & 1.06 $\pm$ 0.11 & 1.75 $\pm$ 0.27 & $<$ 0.32 \\ 
O3E-9 & 289770 & 7577 & 150.0213 & 1.7271 & 3.230 & 222 $\pm$ 5 & 10.11 $\pm$ 0.20 & 1.19 $\pm$ 0.28 & 1.40 $\pm$ 0.26 & 0.86 $\pm$ 0.11 \\
O3E-10 & 278714 & 7597 & 149.9417 & 1.7095 & 3.264 & 136 $\pm$ 4 & 7.28 $\pm$ 0.18 & 1.10 $\pm$ 0.24 & 0.77 $\pm$ 0.23 & $<$ 0.94 \\  \smallskip 
O3E-11 & 274195 & 7604 & 149.9889 & 1.7018 & 3.232 & 262 $\pm$ 5 & 11.37 $\pm$ 0.18 & 2.55 $\pm$ 0.26 & 5.35 $\pm$ 0.26 & 1.22 $\pm$ 0.13 \\  
359521 & 297273 & ---& 150.0043 & 1.7389 & 3.228 & 232 $\pm$ 12 & 3.39  $\pm$ 0.16 & --- &  1.92 $\pm$ 0.26 & $<$ 0.65  \\
361009 & 290562 & --- & 150.0045 & 1.7279 & 3.415 & 201 $\pm$ 25 & 1.49 $\pm$ 0.16 &  0.54 $\pm$ 0.09 &  1.36 $\pm$ 0.12  & $<$ 0.23 \\
361492 & 285414 & --- & 149.9428 & 1.7203 & 3.234 & 169 $\pm$ 15 & 2.29 $\pm$ 0.16 & $<$ 0.36  &   1.15 $\pm$ 0.24 & --- \\ 
363271 & 281091 & --- & 149.9977 & 1.7128 & 3.069 & 203 $\pm$ 7  & 4.60 $\pm$ 0.13 & 1.06 $\pm$ 0.11 &   2.45 $\pm$ 0.20 & $<$ 0.46 \\ 
368172 & 259897 & --- & 149.9322 & 1.6792 & 3.027 & 251 $\pm$ 9 & 8.10 $\pm$  0.26 & 1.89 $\pm$  0.23  &  3.86 $\pm$ 0.26 & 0.71 $\pm$ 0.11 \\
360961 & --- & --- & 150.0077 & 1.7288 & 3.373 & 146 $\pm$ 4  & 6.16 $\pm$ 0.14 & 0.83 $\pm$  0.08 &  1.73 $\pm$ 0.15  & 0.41 $\pm$ 0.08 \\
363778 & --- & --- & 149.9730 & 1.7100 & 3.078 & 134 $\pm$ 11 & 1.56 $\pm$ 0.13 &  0.30 $\pm$ 0.06  & 0.29 $\pm$ 0.08  & ---   \\ \hline
\end{tabular}
\end{center}
{(1) For the \oiii\ emitters, IDs are unique in this paper only. 
For the photo-$z$-selected sources, IDs are extracted from the catalog of \citet{ilbert09}. \\
(2) IDs in the COSMOS2015 catalog \citep{cosmos2015}. \\
(3) IDs in the catalog of \nbk\ emitters from HiZELS \citep{sobral13}. 
We only show numbers here while the IDs given in the catalog are ``HiZELS-COSMOS-NBK-DTC-S12B-**''. \\
(4)(5)(6)(7) Fluxes are shown in the unit of $10^{-17} [\rm erg\ s^{-1}cm^{-2}]$, 
and not corrected for the dust extinction. \\ 
(5) The stellar absorption is not corrected for. \\ 
(6) \oii$\lambda$3726 $+$ \oii$\lambda$3729 fluxes \\
(7) The fluxes with S/N $<$ 3.0 are replaced with the 3$\sigma$ limit values, 
if a line flux is not listed then it was affected by OH skylines. 
} 
\end{table*}%

\begin{table*}
\begin{center}
\caption{
Summary of the estimated physical quantities of our targets.}
\label{table:quantity-summary}
\begin{tabular}{ccccc} \hline {\smallskip}
ID & $\rm \log(M_*)$ &  $\rm A_{FUV}$ & $\rm \log(SFR_{UV})$ & $f_{\rm corr, H\beta}$ \\
 & [$\rm M_\odot$] & [mag] & [$\rm M_\odot yr^{-1}$]  & (1)  \\ \hline
O3E-1 &  9.76 $\pm$ 0.05 & 0.02 $\pm$ 0.19 & 1.17 $\pm$ 0.08 & 1.02 \\
O3E-2  &  9.15 $\pm$ 0.12 & 0.0 $\pm$ 0.33 & 0.82 $\pm$ 0.14 & 1.07 \\
O3E-3 & 10.21 $\pm$  0.15 & 1.06 $\pm$ 0.20 & 1.41 $\pm$ 0.09 & 1.05 \\
O3E-4  & 9.19 $\pm$ 0.14 & 0.74 $\pm$ 0.38 & 0.93 $\pm$ 0.16 & 1.05  \\
O3E-5 & 10.06 $\pm$ 0.15 & 1.06 $\pm$ 0.27 & 1.24 $\pm$ 0.12 & 1.05 \\
O3E-7 &  9.90 $\pm$ 0.06 & 1.00 $\pm$ 0.10 & 1.72 $\pm$ 0.04 & 1.04 \\ 
O3E-8  &  9.88 $\pm$ 0.09 & 0.35 $\pm$ 0.23 & 1.06 $\pm$ 0.10 & 1.08 \\
O3E-9  & 9.07 $\pm$ 0.13 & 0.30 $\pm$ 0.38 & 0.76 $\pm$ 0.16 & 1.03 \\
O3E-10 & 9.23 $\pm$ 0.09 & 0.64 $\pm$ 0.25 & 1.13 $\pm$ 0.10 & 1.06 \\
O3E-11 & 9.86 $\pm$ 0.10 & 1.02 $\pm$ 0.13 & 1.62 $\pm$ 0.06 & 1.06 \\
359521 & 9.44 $\pm$ 0.20 & 0.67 $\pm$ 0.60 & 1.13 $\pm$ 0.24 & 1.14 \\
361009 & 9.81 $\pm$ 0.12 & 1.27 $\pm$ 0.65 & 1.39 $\pm$ 0.26 & 1.11 \\
361492 & 9.13 $\pm$ 0.20 & 0.06 $\pm$ 1.54 & 0.34 $\pm$ 0.62 & 1.06 \\
363271 & 9.71 $\pm$ 0.19 & 1.99 $\pm$ 0.55 & 1.61 $\pm$ 0.22 & 1.05 \\
368172 & 9.82 $\pm$ 0.06 & 1.42 $\pm$ 0.41 & 1.54 $\pm$ 0.17 & 1.04 \\
360961 & 8.46 $\pm$ 0.01 & 1.18 $\pm$ 0.27 & 0.97 $\pm$ 0.11 & 1.00 \\
363778 & 10.48 $\pm$ 0.18 & 0.0 $\pm$ 4.18 & 0.04 $\pm$ 1.68 & 1.21 \\ \hline
\end{tabular}
\end{center}
(1) Correction factor for stellar absorption for \hb. 
The absorption corrected \hb\ fluxes are estimated by multiplying the observed \hb\ fluxes by $f_{\rm corr, H\beta}$. 
\end{table*}

\section{Gas metallicities at the two branches of the KK04 method}
\label{appendix:twometal}

\begin{figure*}[h]
\centering\includegraphics[width=0.5\textwidth]{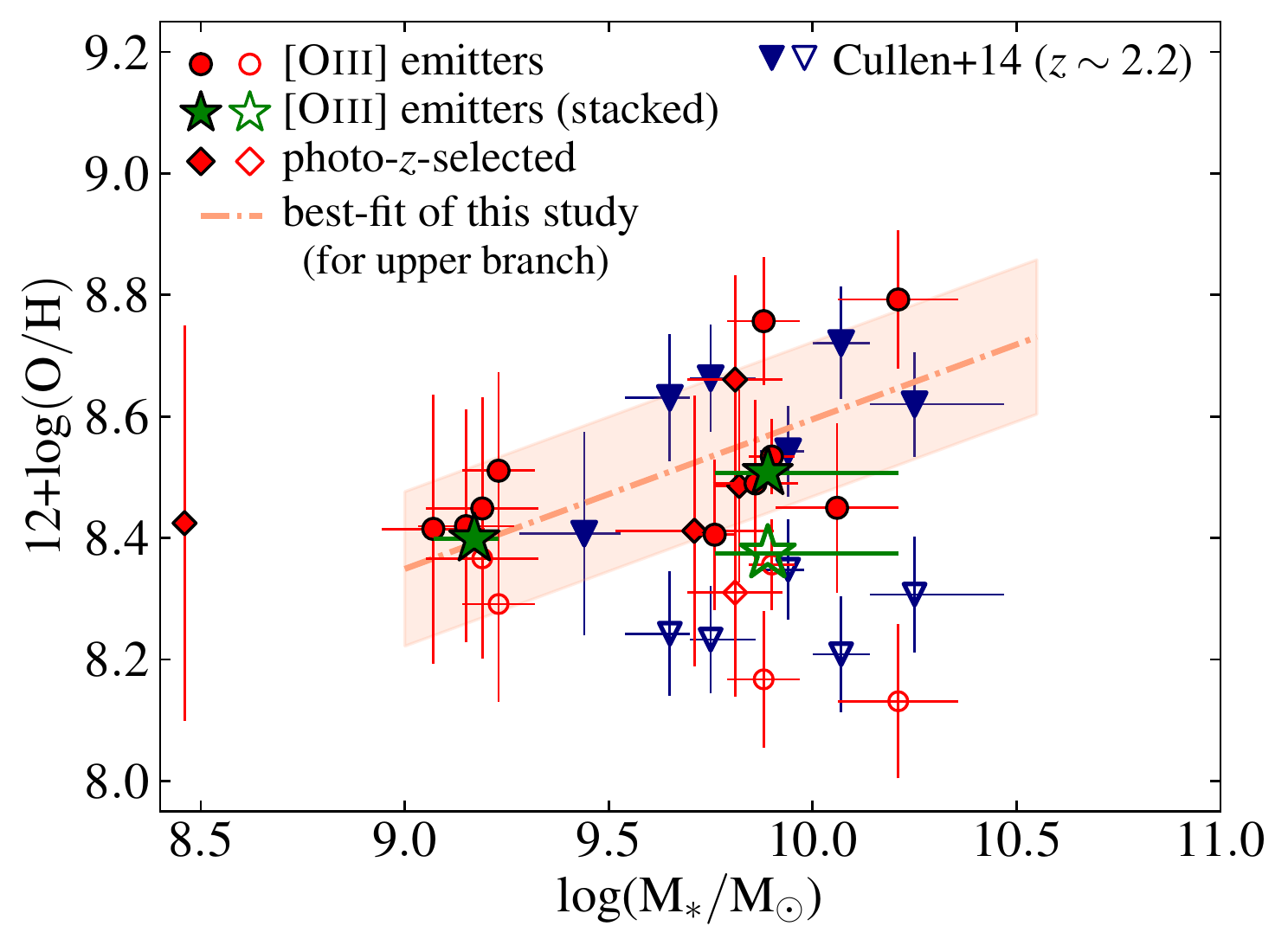}
\caption{
Relation between stellar mass and gas metallicity for our sample at $z\sim3.2$ 
and \citet{cullen14} sample at $z\sim2.2$. 
We show the two solutions obtained by the \citetalias{KK04} method 
(open symbols: lower metallicity branch, filled symbols: upper metallicity branch). 
The dash-dotted line is the best-fitted line derived for the solutions at the upper branch, 
and the shaded region corresponds to $\pm 1\sigma$ errors. 
}
\label{fig:MZ_twobranch}
\end{figure*}

In Figure~\ref{fig:MZ_twobranch}, 
we show the two solutions obtained by the \citetalias{KK04} method  
for our sample and \citet{cullen14} sample. 
Although it is difficult to choose the appropriate branch for our sample with the current data, 
we note that there is no large difference of gas metallicities at a fixed stellar mass 
between our sample at $z\sim3.2$ and \citet{cullen14} sample at $z\sim2.2$ 
when we compare the solutions at the same branch. 
This is consistent with what we see in Figure~\ref{fig:kk04} (a).

%\bibliography{reference}



\end{document}